# NanoVer: An open-source framework for interactive molecular dynamics in extended reality (iMD-XR) on commodity hardware

Mark D. Wonnacott, Luis Ernesto Toledo Castro, Harry J. Stroud, Ludovica Aisa, Mohamed Dhouioui, Rhoslyn Roebuck Williams, Denis Protopopov, Sila Sobrado, David R. Glowacki*

*Intangible Realities Laboratory (IRL), Centro Singular de Investigación en Tecnoloxías Intelixentes (CiTIUS), University of Santiago de Compostela, 15782, Santiago de Compostela, Spain*

**drglowacki@gmail.com*

## Abstract

This article outlines 'NanoVer', an open-source software framework which enables groups of people to co-habit the same virtual space and manipulate real-time MD (Molecular Dynamics) simulations of flexible 3D molecular structures with atomic-level precision as if they were tangible objects, an approach that we call '**i**nteractive **M**olecular **D**ynamics in e**X**tended **R**eality' (**iMD-XR**). Distinct from our earlier iMD work that relied on tethered PC-VR systems with large graphics cards, NanoVer represents a change in philosophy, emphasizing compatibility with standalone mobile consumer XR hardware and corresponding software APIs. The NanoVer architecture enables multiple XR clients and/or Python clients to simultaneously communicate with a flexible server architecture that can carry out a range of tasks, including for example: recording iMD-XR sessions, static structure visualization, and MD trajectory visualization. NanoVer allows researchers, educators, and students to fluidly move between AR and VR environments, to explore creative new approaches to molecular research and education, including for example: molecular conformational sampling, protein-ligand binding, molecular psychophysics, training AI agents to sample molecular transitions, and a new interface which allows iMD-XR participants to sketch 3D conformational paths which automated agents can then follow. As an immersive platform that offers new ways to understand, engineer, communicate, and interact with dynamical behaviour at the nanoscale, NanoVer invites us to imagine new ways for combining human intelligence (e.g., spatial cognition and design reasoning) with machine intelligence. To expand NanoVer's accessibility, we have published a version to the Meta Horizon Store, for easy download by those with a Meta Quest 3/3S headset, to explore pre-recorded iMD-XR trajectory visualizations and set up their own multi-user system.

## 1. Introduction

Human cognition has co-evolved alongside the seemingly inexhaustible sensory richness of natural environments whose time-evolving structures, sounds, textures, and sensations, we navigate using the interfaces afforded by our sensory systems, including vision, hearing, touch, smell, taste, balance, proprioception, interoception, nociception, thermoception, etc. The last few years have seen an explosion of investment in computational power, with enormous data centers now supporting the training of hyperdimensional large language models (LLMs) with hundreds of billions (or even trillions) of parameters. Despite this vast dimensionality, the interfaces we use to engage with computational models like LLMs are surprisingly limited: mostly 1D strings of alphanumeric language tokens input on keypads and 2D screens. Such limited interfaces contract the vast potential and richness of our human sensory experience into one or two spatial dimensions, which some scholars have suggested represents an *enslavement* of our human potential to the demands of computational machinery.(*1-3*) Our tendency to contract our potential for sensation and expression into the relatively limited interfaces demanded by machines has been highlighted since the early days of human computer interaction (HCI), with pioneers like Myron Krueger observing that "human-machine interaction is usually limited to a seated (person) poking at a machine with (their) fingers or perhaps waving (their) hands over a data tablet." (*4*)

During the past several years, advances in computer vision, mobile graphics, motion sensing, micro-processor engineering, and GPU-accelerated computation have created new possibilities for using "extended reality" (XR) approaches including augmented reality (AR) and virtual reality (VR) to visualize, understand, and interact with complex 3D data using multiple sensory modalities. These developments have enabled us to undertake a research program focused on *expanding the richness of the human perceptual experience* whilst engaged with computational science workflows. In contrast to many of the emerging computational tools for molecular science – which are designed to achieve efficiency and automation that bypasses the need for a human in the process of scientific insight and discovery – we have instead taken a human-centric approach focused on enhancing the richness of human experience whilst engaging with the beauty of nature's dynamical structure at the nanoscale.(*5*) Our primary application domain has been on molecular physics simulations, a relatively mature domain of scientific computing which has been recognized by various Nobel Prizes and has now become essential to research and education in the nanosciences. To the extent that our research has focused on developing a richer human perceptual experience, it implicitly recognizes the distinctly human "pleasure of finding things out" noted by the late Richard Feynmann.(*6*) As most of us probably know from our own first-hand subjective experience, multi-sensory experiences tend to be more cognitively poignant than single-sensory experiences – i.e., we tend to have stronger memories of experiences which involve multiple senses,(*7*) and recent research shows that multi-sensory experiences enhance learning.(*8, 9*) Such emphasis on human experience recognizes the importance of human subjectivity, a research domain that is expanding broader framing within the growing field of 'Consciousness Science',(*10, 11*) but which sometimes seems at odds with classical scientific emphases on 'objectivity'.(*12*)

Nevertheless, research on subjective experience seems destined to grow, in part to understand what distinguishes human experience, intelligence, feeling, and emotion from machine intelligence.

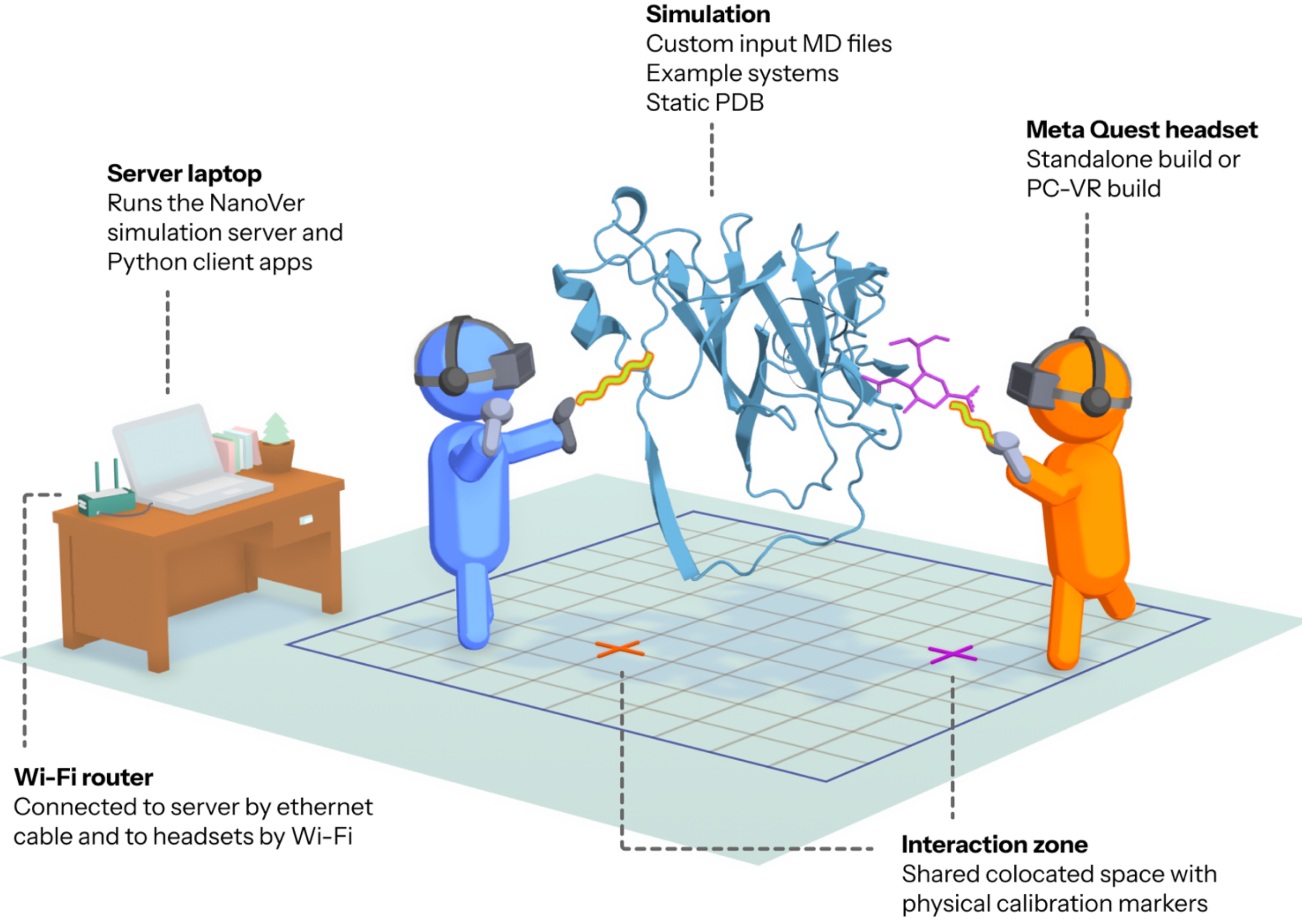

*Fig. 1: Schematic showing the NanoVer iMD-XR framework for two participants that are colocated (i.e., in the same physical space). Each participant wears a standalone HMD and holds handheld wireless controllers, which they use to manipulate a real-time MD simulation of Neuraminidase. Each participant's perspective of the molecular system is rendered using the HMD's onboard mobile graphics processors. MD calculations and management of global user position data take place on a local NanoVer server computer, shown here as a laptop connected via ethernet cable to a Wi-Fi router. The figure shows a setup where the MD server and HMD clients communicate via a local area network (LAN). Additional python client apps run in parallel on the NanoVer server laptop.*

This paper outlines *NanoVer*, an open-source iMD-XR software framework we have been developing over the last few years (illustrated in Fig 1), which enables multiple people wearing head-mounted displays (HMDs) to co-habit the same immersive environment and interactively manipulate rigorous molecular physics simulations running in real-time, as if they were tangible objects. Beyond its primary iMD-XR functionality, *NanoVer* also enables the visualization of static structures and pre-recorded trajectory files. *NanoVer* takes its name from the words *Nano* and *Ver* (which means 'to see' in Spanish, Portuguese, and Galician). Whereas most molecular simulation workflows are restricted to flat 2D displays and rely either on a mouse or a keyboard, iMD-XR enables the exploration of real-time molecular simulations in both 3D and 4D, incorporating senses beyond vision including sound(*13*) and proprioception.(*14*) Beyond the technical challenges required to implement such approaches, they also require a degree of *aesthetic imagination* – for the simple reason that the world of atoms and molecules (along with their corresponding physics) is not accessible in our day-to-day experience of the world. This stands in contrast to applications of XR technologies in other domains like surgery,(*15*) where there is an unambiguous 'real-world' design reference for how the simulator should 'look' and 'feel' – that is, one can carry out detailed comparisons assessing whether the simulator has fidelity to the 'real-world' experience of their human senses.

NanoVer is one amongst several XR-compatible molecular science software tools which have been presented in the literature over the last several years (an exhaustive list of which is beyond the scope of this particular article); however, several excellent reviews,(*16-26*) software papers,(*27-40*), and applications papers (*18, 38, 41-47*) are available for the interested reader. NanoVer builds on our previous work developing an open-source interactive molecular dynamics software framework called 'Narupa', which was described by O'Connor et al.(*48*) and Jamieson-Binnie et al.,(*49*) and has since been applied in a range of different research contexts, including for example:

- Interactive ligand-protein binding studies for a range of systems including trypsin, neuraminidase, and HIV-1 protease(*50*)
- Docking of inhibitors to the main protease of the SARS-CoV-2 virus(*51, 52*)
- Calculating free energy along interactively sampled protein-ligand binding pathways(*53*)
- Using real-time density functional theory to train neural nets to learn reactive potential energy surfaces(*54, 55*)
- "Gamifying" the discovery of chemical kinetics reaction networks(*56*)
- Investigating atomic and molecular transport dynamics in solid state materials like zeolites(*57*)
- Using established methods from psychophysics to study so-called pseudo-haptic sensations that arise whilst participants 'feel' the properties of different molecular simulations in XR(*14, 58*)
- Developing new strategies for teaching university students about enzyme catalysis(*59*) and molecular interactions (*60*)

NanoVer has been designed to run on the latest generation of 'standalone' HMDs, including the Meta Quest 3 and 3S. NanoVer thus represents a significant step forward compared to Narupa, which was designed around the prior state-of-the-art: PC-VR systems with HMDs connected via graphics cables to gaming PCs equipped with high-end GPUs, and 'outside-in' tracking of HMD position/orientation using externally placed optical 'base-stations' which emit structured infrared light detected by sensors on the HMD. Standalone HMDs by comparison use mobile graphics processors and 'inside-out' optical tracking (i.e., onboard cameras calculate the HMD's position and orientation relative to the physical world). At the time of writing this article, the most recent versions of the Meta Quest (3 and 3S) cost between \$400 – \$600. Unlike PC-VR systems, standalone systems do not require expensive PCs equipped with a high-end GPU. Standalone mobile graphics processors are unable to match the graphics performance available with a high-end PC GPU; however, for the vast majority of researchers and educators standalone systems offer a more affordable solution with easier setup and vastly more portability.

NanoVer's compatibility with the recent OpenXR standard means that it runs on all major HMDs and enables access to an expanded feature set, including the ability to navigate the reality-virtuality continuum shown in Fig 2. As mentioned in the abstract of this article, XR technologies include both AR and VR; however, NanoVer does not take a binary either/or approach in regard to AR/VR. Instead, NanoVer recognizes that AR and VR are two points along a broader Reality-Virtuality Continuum (illustrated in Fig 2),(*61*) which we have described as a 'reality dial' in the context of recent neuroscience research studies. The 'reality dial' approach treats virtuality and augmentation like dimmer switches, which can be varied continuously to achieve different overlays of the 'virtual' onto the 'real'.(*62, 63*) The 'reality dial' approach is possible via the 'passthrough' functionality of the onboard cameras embedded in standalone HMDs, which enable the user to see the external world through the HMD.

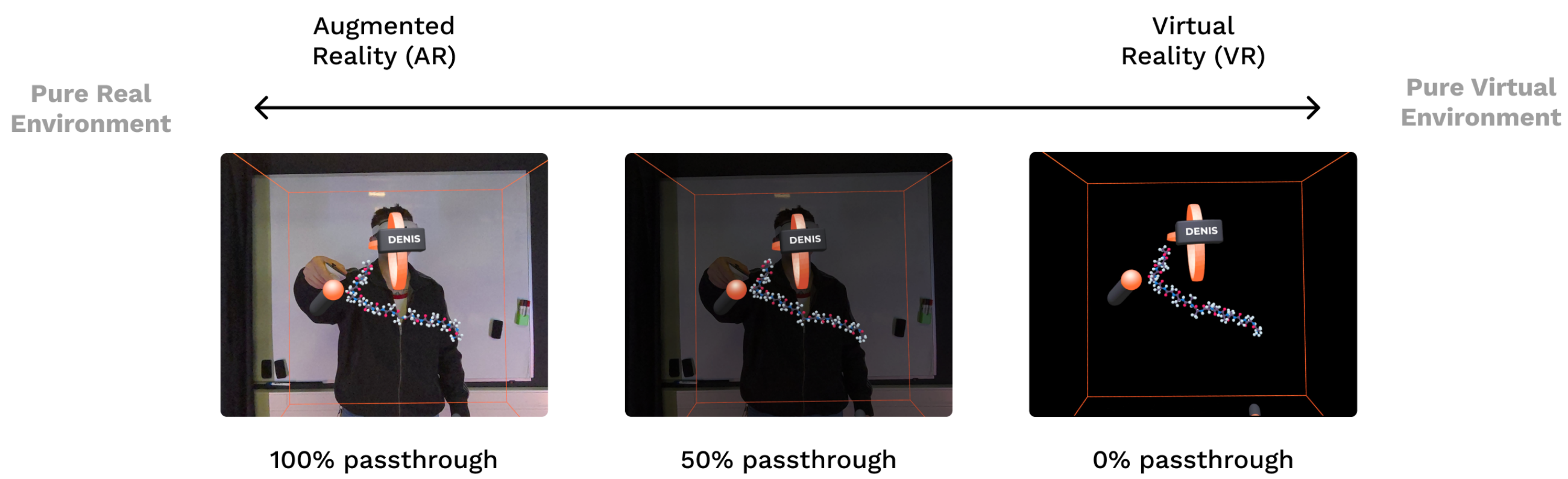

Fig. 2: Schematic of the 'Reality-Virtuality Continuum', which illustrates how much of the 'real environment' is mixed into the digital display. VR is located at one side of the continuum and AR is located toward the other side. NanoVer allows users to set the 'reality dial' to any value they like, effectively occupying any point along the continuum.

NanoVer allows users to set the value of the 'reality dial' by choosing how much 'passthrough' they see – i.e., how much of the real environment is mixed into the virtual display. Low values of the reality dial correspond to an AR display, whereas high values correspond to a VR display. The flexibility offered by the reality dial significantly improves the ergonomics of NanoVer compared to Narupa (which was limited to VR and therefore restricted to a single point on the RV Continuum). The same onboard cameras that enable the 'reality dial' also enable optical hand tracking, so that HMD-wearing participants like those shown in Fig 1 can use their hands (rather than controllers) for interacting with and navigating the simulation.

In what follows, we introduce the NanoVer software and describe its architecture, which is comprised of server-side applications (*27*) and client-side applications that communicate via a defined protocol. To the best of our knowledge, NanoVer is the only open-source multi-person iMD-XR software package available on commodity devices like the Meta Quest 3 and 3S. NanoVer has been designed with flexibility in mind, given its primary use as a tool to support new approaches toward research and education, which are often emergent and not necessarily well-defined from the start. This article describes the NanoVer architecture, provides some guidance on setting up a multi-person NanoVer iMD-XR environment, describes studies to benchmark NanoVer's performance, and illustrates NanoVer's flexibility by discussing various ongoing research applications. Alongside the preparation of this article, we have recently published a version of NanoVer on the Meta Horizon Store, making it easy for anybody with a Meta Quest 3 to quickly download NanoVer and visualize a sample of pre-recorded iMD-XR trajectories (https://www.meta.com/experiences/nanover-imd-xr/33606061302340842/). The Supporting Information contains further details on the version of NanoVer available on the Meta Horizon store.

## 2. NanoVer

### *2.1 Overview*

Over the last several years, various open-source source tools have emerged within the computational chemistry ecosystem, supporting a range of tasks that include molecular visualization,(*64-66*) molecular dynamics(*67, 68*), analysis of molecular simulations,(*69, 70*) and file interoperability.(*71, 72*) These tools can be combined to support bespoke workflows within research and education. NanoVer is designed to sit within this stack, building on these established frameworks while providing real-time, collaborative interaction in XR for human-in-the-loop exploration. The manner in which iMD-XR users can bias simulations in real-time has been previously described by O'Connor et al.(*73*) and therefore is not repeated here. The various components of the NanoVer software framework can be understood with reference to Fig 1, which illustrates the physical setup for a multi-person NanoVer session in which users are 'colocated' (i.e., co-present within the same physical and virtual space). In its simplest setup, NanoVer includes the following:

1. *iMD-XR client applications*. In Fig 1, multiple participants wear passthrough-capable XR headsets, each running the iMD-XR client software. Upon entering the XR environment, each participant proceeds to calibrate their space, aligning their location in the virtual space with their physical location from the perspective of other participants. It is also possible for participants to join remotely (i.e., from different physical spaces), in which case calibration is not required.
2. *The NanoVer server application*.(*27*) In Fig 1, the NanoVer server runs on a single high-performance computer which is equipped with adequate RAM, CPU power, and GPU power to run MD simulations. The NanoVer server uses Python to wrap MD simulation engines so that they are accessible to the networked XR clients, providing them access to live system data (e.g., topology, atomic positions, etc.) and the ability to introduce biasing forces in real time.
3. *Server-Client communication.* In Fig 1, the server computer is connected by ethernet cable to a high-quality router that provides Wi-Fi connectivity to the XR headsets. Data is exchanged between server and clients over this network using the standard WebSocket protocol. Connectivity of remote clients can be supported by exposing the server directly over the internet or via VPN.
4. *Python/Jupyter client applications.* NanoVer's flexibility arises from the fact that it exposes the MD simulation through a networked client–server architecture whose interface can be accessed via modular

Python scripts or Jupyter notebooks that a user can easily adapt and extend to suit their needs. This makes NanoVer simulations accessible to a suite of utilities that support common computational workflows, including: preparing the simulation systems (topologies, forcefields, etc.); controlling the simulation and visualization parameters during a real-time iMD-XR session; using NGLView for real-time visualization of an iMD simulation;(*65*) conversion between NanoVer trajectory data (described below) and MDAnalysis;(*69, 74*) and the ability to trigger the execution of notebook code from within the XR client. In Fig 1, the Python/Jupyter applications run on the same computer as the NanoVer server.

NanoVer treats iMD sessions as first-class procedures whose trajectories and steering inputs can be logged, visualized *post facto*, re-simulated, and analyzed using toolchains similar to those for conventional non-interactive MD runs. NanoVer provides an analogue to the usual trajectory recording which (in addition to the usual data stored in an MD trajectory) also includes information about the live interactive forces that have been introduced, as well as the session-based data of human participants within the XR space: e.g., their movements with respect to one another and also the molecular system; their head and hand movements as they inspect and manipulate the molecule; and other bespoke data introduced to support a particular experiment or workflow. As such, NanoVer can be used to visualize real-time iMD simulations, as well as replays of iMD-XR recordings, allowing multiple users to e.g., review together a previous experiment. NanoVer can also be used to collaboratively visualize existing trajectory files or static structures within multi-user XR.

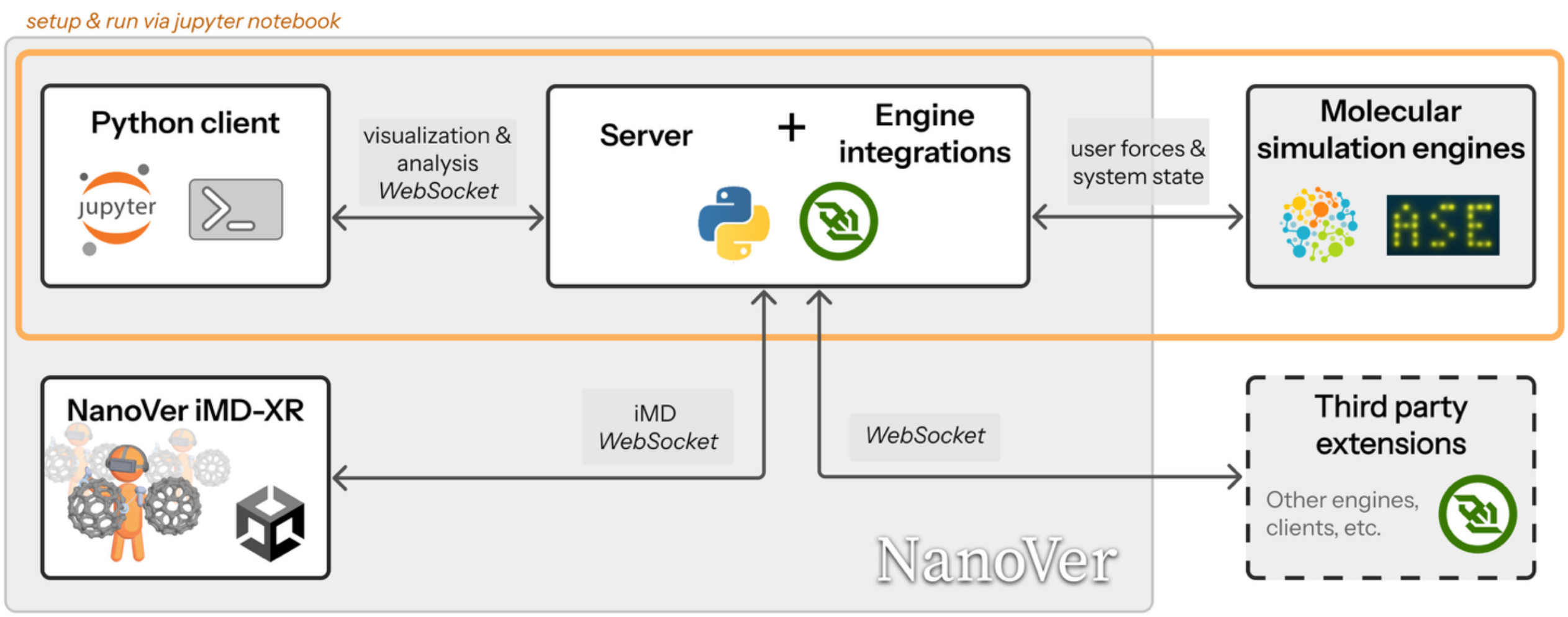

*Fig. 3: Schematic of relationship between the various NanoVer software components (which are shown within the grey box) and third party software packages.*

### *2.2 Architecture & repository structure*

The NanoVer iMD-XR setup is comprised of a number of interconnecting software components which are schematized in Fig 3. The molecular simulation engines such as OpenMM and ASE are typically instantiated on the same machine as the server, communicating directly in Python. Python and Jupyter clients are often run on the same machine as the server (as in Fig 1) or on separate machines, depending on the specifics of the setup and the computational cost of the MD simulation. iMD-XR headset clients and third-party integrations are typically run on other machines communicating with the server over the network using the WebSocket protocol.

The WebSocket-based network communication in NanoVer is structured around three core message types: 'frame' updates, 'state' updates, and 'command' requests and responses. Frames in NanoVer are individual snapshots (primarily updated atom positions) of the simulated system, broadcast frequently enough to visualise the simulation in real time. State represents all of the non-MD-simulation data used to provide a synchronized experience: e.g., the specific parts of the molecular system with which the user wants to interact, or the visualization configuration of a particular molecular system. In a multi-person context, the state also communicates the positions of other users and their interactions. Commands are client-triggered events that can be used for a range of different purposes, including for example controlling playback of the simulation (play/pause/reset), switching between simulations, and custom-defined operations a researcher may want to implement (e.g., saving simulation snapshots of a system). The software components of NanoVer shown in Fig 3 are available in two open-source repositories: the so-called 'NanoVer server' repo and the NanoVer iMD-XR client repo.

The NanoVer server repo (github.com/IRL2/nanover-server-py) includes the following:

1. Python code which: (a) enables the client-applied interactive forces to be integrated into real-time simulation engines and (b) extracts from the running simulation the information to be shared over the network (primarily atom positions, but also forces, velocities, energy, and temperature if desired). At present NanoVer provides integrations with ASE(*68*) and also the GPU-accelerated MD engine OpenMM.(*67*) NanoVer users can develop their own bespoke integrations to other MD engines either in Python according to NanoVer's modular interface (e.g., a forthcoming LAMMPS integration(*75*)), or by implementing the simple network protocol to connect a force engine as a client to the server (https://irl2.github.io/nanover-docs/concepts/base-protocol.html)
2. A Python server, which manages the exchange of frame data, state updates, and control commands between the simulation engine integrations and the various clients.
3. Code which enables Python clients to connect to the server, access the raw simulation data, and access an expanded set of simulation controls beyond those which are available in the user interface of the iMD-XR client (for example enabling users to make selections, change the molecular visualization, plot real-time data, automate interactions and restraints programmatically or with AI agents, etc.).
4. A set of Jupyter notebook examples and tutorials that streamline the workflow required to set up and connect the various Python scripts which are generally used during simulation – e.g., setup and loading of the initial molecular systems, starting the simulation engines and connecting them to the message-passing server, and performing live data plotting and analysis of the interventions as they are made from the XR headset clients.

The iMD-XR client repo (github.com/IRL2/nanover-imd-xr) includes sources for NanoVer iMD-XR, our headset-based client, along with routines that enable the XR headset clients to connect to the server. The iMD-XR client provides an interface for: (a) visualizing either a live simulation or playback of a pre-recorded session; (b) interactively applying forces using the XR controllers, and (c) triggering the various control functions such as play/pause/reset from in-world user interfaces. Given the complexity of developing intuitive user interfaces within XR, the control command interface within the iMD-XR client tends to be only a subset of that which is available via the Python client – i.e., control commands tend to be migrated to iMD-XR in cases where there is a clear need given their frequency of usage. The headset client is developed in the latest version of Unity (Unity Engine 6 at time of writing) using the modern Meta XR SDKs for XR support,(*76*) Nerdbank.MessagePack(*77*) for serialization, and NativeWebSocket(*78*) for connectivity.

For transporting messages between clients, NanoVer utilizes WebSocket for establishing network connections between different instances of the software, and MessagePack (*79*) for efficient serialization of message data. WebSocket has been chosen for its near-universal support across platforms, simple message-based interface, as well as its unique ability to interoperate with browser technologies such as WebXR. MessagePack is chosen for its mature third-party support across many platforms (particularly in Unity, Python, and the browser). Moreover, its schemaless nature is well suited to the fact that NanoVer is designed as a flexible research framework which can support diverse projects with data schemas that are often not well established in advance. NanoVer's adoption of WebSocket also ensures an easy transition to WebTransport(*80*) which expands WebSocket's capabilities for high-performance real-time communication.(*81*) The move to WebSocket and MessagePack represents an evolution of NanoVer compared to its predecessor Narupa, which respectively utilized gRPC (https://grpc.io/) for network communications and Protobuf for data serialization. Given that WebSocket and MessagePack represent two mature, well-supported, and cross-compatible off-the-shelf technologies, their use within NanoVer minimizes the difficulty for other interested researchers to develop software that interoperates with NanoVer, which we previously found to be a point of friction with the gRPC-based implementation.

### *2.3 Visualization*

The NanoVer iMD-XR client supports a selection of standard molecular visualization techniques (illustrated in Fig 4) for visualizing real-time simulations, trajectory files, or static structures. (*82*) We use the raycasting impostors technique for high-performance rendering of the smooth shapes such as balls, cylinders, and ribbons common to molecular visualization.(*49, 83*) Beyond the ball-and-stick renderers which we used to generate the performance benchmarking results in section 3, we highlight two particular renderers that are available in NanoVer: an implementation of 'Hyperballs',(*84*) which is particularly well suited to visualizing bond dynamics (especially the intermediate stages of bond making and breaking), and a ribbon-based "cartoon" renderer for proteins that morphs smoothly as the real-time protein secondary structure changes with deformation.(*83*) These renderers have been updated and optimized to function correctly within the additional limitations of the mobile GPU used by the standalone Meta Quest headsets.

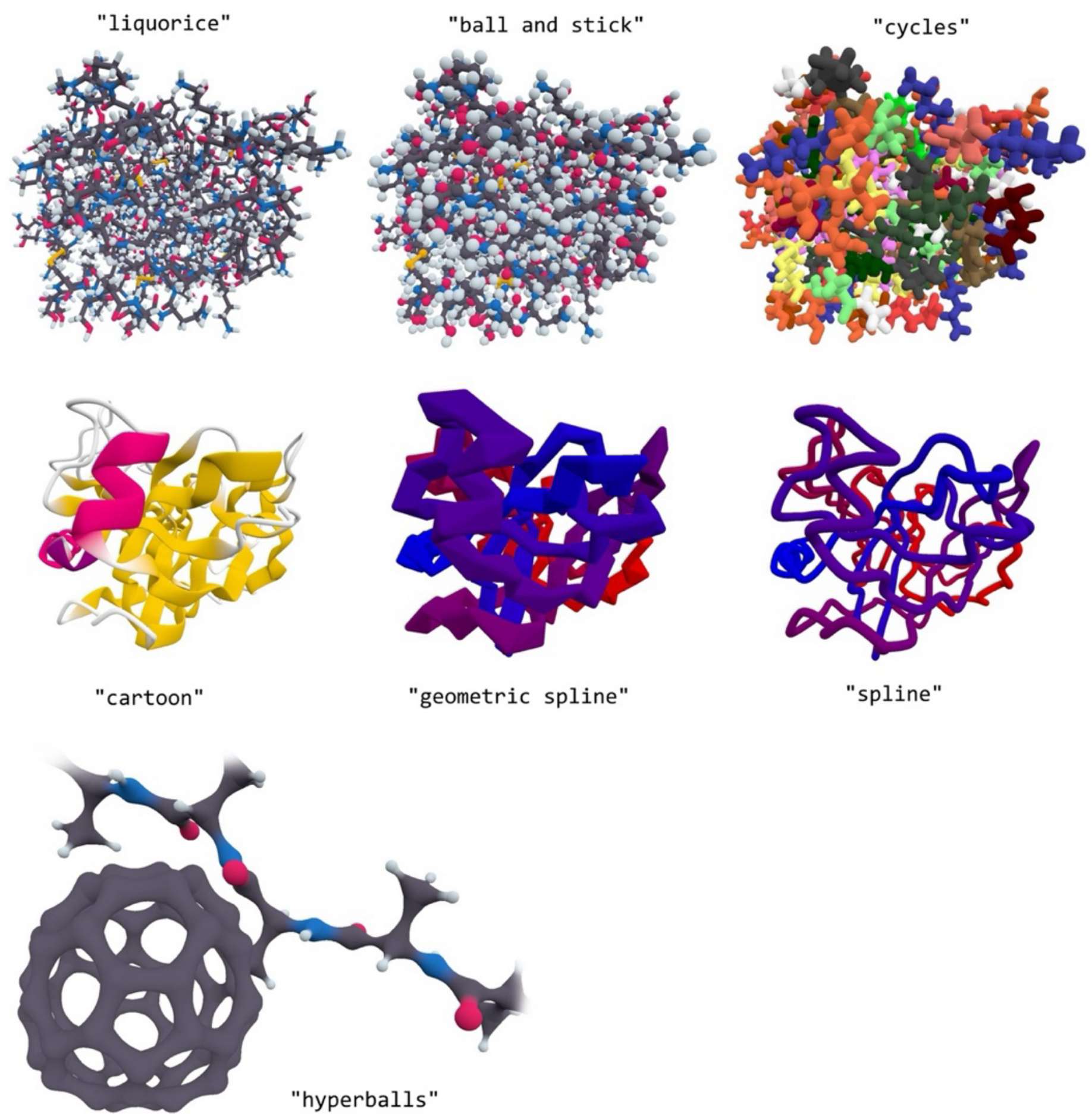

Fig. 4: Illustration of the most commonly used renderers currently available within NanoVer

### *2.4 Setting Up & Running NanoVer*

Setting up NanoVer requires installation of both client and server software, with detailed instructions available at irl2.github.io/nanover-docs/installation.html. The NanoVer client can be run and installed on your Meta Quest 3 HMD either via the Meta Horizon store (search for 'NanoVer') or else by enabling 'developer mode' and sideloading the GitHub releases which are available at github.com/IRL2/nanover-imd-xr/releases (both standalone android and PC-VR builds). To run iMD-XR using the version of NanoVer on the Meta Horizon store requires having a local server to which you can connect (see instructions in the Supporting Information); otherwise you will only be able to visualize pre-recorded trajectories. The nanover-server conda package is available at https://anaconda.org/irl/nanover-server with some form of conda (e.g. miniforge). For scientific workflows, we typically launch the MD simulations from Jupyter notebooks, which enables us to set up the molecular systems and their corresponding forcefields, along with the desired visualization parameters. Once the simulation is running, client apps can connect to the server over the local network. Further details are available in the documentation at irl2.github.io/nanover-docs/tutorials/tutorials.html. For colocated multi-user setups as illustrated in Fig 1, in-app calibration (with instructions) ensures that the space is correctly oriented for all participants.

*2.4.1 Institutional Issues*

*Eduroam* and other institutional networks often enforce security policies that block direct device-to-device communication required by NanoVer and Meta Quest in general. To avoid these difficulties, our lab has an autonomous private network that does not restrict communication between the NanoVer software, the XR headsets, and other machines. If you cannot negotiate a subnetwork with your institution, an isolated Wi-Fi router without internet access also works (as shown in Fig 1). We have also used ad-hoc mobile hotspots successfully, but these do not perform as well as a dedicated router (see Fig 5). To develop customized versions of NanoVer or load customized *.apk executables onto the HMD requires 'developer access', which is sometimes incompatible with institutional device management policies, and therefore requires IT manager support.

*2.4.2 Documentation & Examples*

Full documentation of NanoVer's messaging protocol for publishing and receiving live simulation frames, shared state updates, and for defining and executing commands is available at https://irl2.github.io/nanover-docs/concepts/base-protocol.html. Moreover, the NanoVer documentation at irl2.github.io/nanover-docs/index.html illustrates various aspects of typical workflows, including for example:

- Running a previously published iMD-XR application of Tamiflu bound to Neuraminidase(*50*) which involves loading AMBER input files into the OpenMM simulation engine, with various customizations of the protein-ligand visual representation.(*85*)
- Loading a PDB input file of LSD bound to the 5-HT2B human serotonin receptor(*86*) as a static structure via MDAnalysis which is then served through NanoVer for visualization in the XR client.(*87*)
- Serving multiple systems with different preset visualizations and switching between them within the HMD.(*88*)
- Recording NanoVer iMD-XR sessions into standard trajectory file formats.(*89*)
- Using MDAnalysis to analyse data from NanoVer recordings.(*90*)
- Defining custom commands which can be triggered from within the HMD as part of a custom workflow.(*91*)
- Using WebSocket + MessagePack outside of NanoVer to communicate with a NanoVer server and extract frame data.(*92*)

## 3. Performance

For non-interactive MD, the general aim is for the simulation to run as fast as possible on any given hardware. Given that the aim of iMD is to allow the user to meaningfully guide a system in real-time, the dynamics must evolve at a timescale which a human can interpret and manipulate if they wish. As such, iMD is subject to similar considerations as a multi-person video game – i.e., the simulation must progress at a fixed rate. For small systems, this means that NanoVer will underutilize the available compute, whereas for systems above a certain size it will be unable to sustain the real-time simulation update rates required for iMD. The perceived quality and usability of iMD-XR depends on sustaining a smooth and responsive interaction experience in which the user can see the system evolving smoothly and make manipulations that appear to take effect continuously and in real-time without significant delay while sustaining a smooth XR experience. To assess the performance and quality of the NanoVer experience, we have utilized three real-time performance metrics:

(1) *Frame updates per second (FUPS)*. The NanoVer server sends position updates to clients at a fixed rate; if those updates cannot be delivered at that rate, the system becomes overloaded and the sense of smooth motion is lost, creating a scenario where the user does not reliably see the current state of the simulated system.

(2) *Round trip time (RTT)*. This measures the time that elapses between (a) the instant that a client requests a change on the server (e.g. a user exerts a force to manipulate an atom position, or updates one's own avatar position) and (b) perceiving the result of that change. Keeping this time below a certain threshold ensures that one's own interactions, as well as the movements of other users, appear to take place in real-time without any noticeable delay.

(3) *XR FPS (frames per second)*. This measures the number of graphics frames rendered per second on the XR client. The XR FPS determines how responsive the XR system is to changes in a user's head movements, and must be maintained above a certain threshold so the user doesn't suffer cybersickness.(*93*)

We have identified a number of factors that affect these performance characteristics, including:

(1) *Network capacity.* As shown in Fig 1, the network that connects the headsets to the server consists of a router and a combination of Wi-Fi and/or cabled ethernet connections. The precise network configuration has a maximum data throughput capacity. The lower this capacity, the fewer atom position updates can be sent per second, which implies smaller simulations and fewer simultaneous users.

(2) *Wi-Fi quality*. The quality of a Wi-Fi connection can degrade with distance, interference, number of users, and load. Lower quality entails reduced data throughput capacity, along with increased and more erratic RTTs.

(3) *System size*. As the number of atoms in a simulation increases, so too does: (i) the computational cost of the simulation itself; (ii) the quantity of data to transfer over the network; and (iii) the rendering cost to visualize the full system. These factors impose limits on the max system size that can be treated using iMD-XR.

(4) *Hardware compute power*: Typically, the more powerful the CPU/GPU in the machine that runs the server, the faster an MD simulation can run, and therefore the larger the possible system size. On the HMDs, more GPU power enables larger systems to be rendered while maintaining the required framerate.

(5) *User count*. With $n$ users in the setup, the atom position updates must be transmitted $n$ times. Each user puts an equal load on the network capacity, so that doubling the number of users halves the available capacity and therefore halves the number of atoms which can be simulated using iMD-XR.

To display smooth motion, NanoVer sends at 30 FUPS. The size of each update is dominated by the atom positions. Each atom's cartesian position is defined by three 32-bit floating point numbers (96 bits in total). A single atomic position updated at 30 FUPS therefore requires 2,880 bits per second of network capacity; one thousand positions at 30 FUPS requires 2.88 Mbps of network capacity. For a multi-person room-scale XR experience, Wi-Fi is generally preferable for practical and safety reasons, in which case the headsets are connected to the network via Wi-Fi. Because simulation updates must first travel from the server to the router and then from the router to the XR HMD clients, connecting the server to the router via ethernet cable (as illustrated in Fig 1) minimizes the load on the Wi-Fi channel.

In what follows, we present a number of tests aimed at characterizing the performance of NanoVer. These tests utilize the setup illustrated in Fig 1, with the following hardware:

(1) *Server Laptop:* A Legion Pro 5 16IRX9 laptop (with Nvidia GeForce RTX 4070 Laptop GPU) which is capable of sustaining 30 FUPS for systems up to 180,000 atoms. Tested in conjunction with the router, it has a max Wi-Fi throughput of 893.69 Mbps, and a max cabled ethernet throughput of 944.57 Mbps.

(2) *Meta Quest 3 XR* headsets running the clients, which our benchmarking tests indicate has a max Wi-Fi throughput of 758.24 Mbps. To carry out the PC-VR rendering tests, we connected the headsets to an Alienware Gaming Laptop (AW15R3-7003SLV-PUS) with GeForce GTX 1070.

(3) *Wi-Fi router:* An ASUS TUF Gaming AX6000 Wireless Router, providing Wi-Fi 6 5GHz with 80MHz channel width. This model has a theoretical max throughput of 1201 Mbps; our benchmarking tests showed an upper bound on Wi-Fi throughput of 893.69 Mbps (in conjunction with *Server Laptop*) in a single client case.

Based on these benchmarks, the network bottlenecks are $893.69/n$ Mbps for the router (where $n$ is the number of users), and 758.24 Mbps for the Meta Quest 3 (which is independent of the number of users). From these values, we can calculate a theoretical limit of 180k atoms for one user (i.e., the max number of atoms that can be simulated at 30 FUPS), 155k atoms for two users, 103k atoms for three users, and 78k atoms for four users. When the server is not connected to the Wi-Fi router via ethernet cable and instead shares Wi-Fi with the clients, the data transfer load doubles, giving a limit of 155k atoms for one user, 78k atoms for two users, 52k atoms for three users, and 39k atoms for four users.

Figs 5 and 6 show performance measurements of the actual setup as a function of the number of atoms in the simulated system, reported as the rate of frame updates received by the client (corresponding plots of RTTs are available in the Supporting Information) for the case of 5 MD simulation steps per visualization update. Dropping below 30 FUPS signals that the setup is overloaded and unable to sustain usable, responsive iMD-XR. The extent of the drop gives an indication of how severely overloaded the setup is. Fig 5 illustrates the performance benefits of connecting the Wi-Fi router to the server laptop via ethernet cable compared to a wireless connection or mobile hotspot. Fig 6 shows that NanoVer performance drops as the number of users increases, but can nevertheless sustain iMD-XR simulations for systems of 75k atoms for four-user setups. 75k represents a larger system size than we currently work with, given the limitations on rendering.

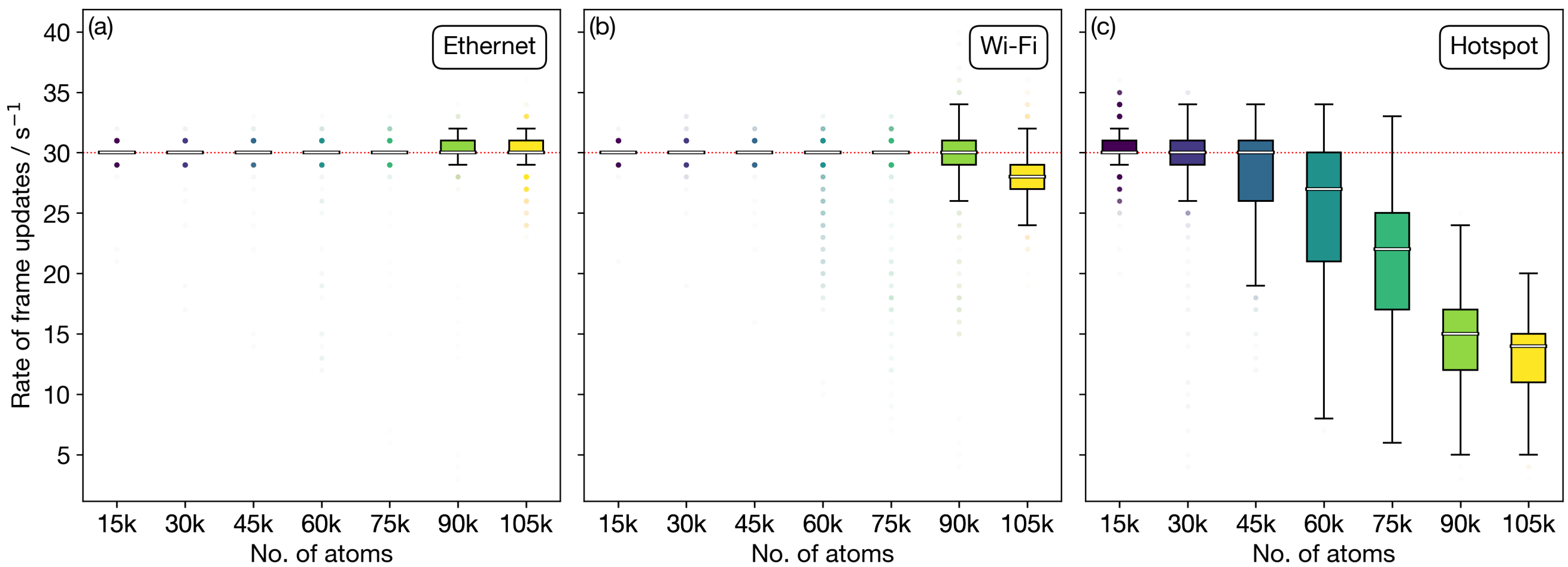

Fig. 5: NanoVer network performance metrics for one client connected to the server. The plot shows performance as a function of the number of atoms in the simulation for a setup where the NanoVer server laptop and clients are connected via (a) ethernet cable; (b) Wi-Fi router and (c) mobile hotspot. The box extends from the first quartile (Q1) to the third quartile (Q3) of the data, with a line at the median. The whiskers extend from the box to the farthest data point lying within 1.5x the inter-quartile range (IQR) from the box. Data points extending past the end of the whiskers are explicitly shown.

Fig 7 shows measurements of the headset rendering performance as a function of the number of atoms in the visualized system, reported as XR FPS. Modern XR rendering utilizes a combination of approaches that dynamically scale the rendering quality to sustain a framerate, which can sometimes cause framerates to fluctuate.(*94*) The baseline for high quality experiences is a steady 72 FPS, but we find 36 FPS to be acceptable for typical scientific use of NanoVer. To give smooth details, NanoVer renders by raycasting implicit surfaces,(*49, 83*) which trades a large vertex count for a heavier shader load, and in turn means that the apparent size of atoms on-screen can affect performance just as much as the quantity of atoms in the scene. For this reason, we tested NanoVer rendering performance in three typical user points of view (POVs) relative to the position and size of the simulation ('Desktop', 'External', and 'Immersed' in Fig 7), where the system being rendered was obtained from real-time simulations of TIP3P water boxes with varying molecule counts. Moreover, we also compared rendering performance using mobile graphics processors ('Standalone') versus a laptop GPU ('PC-VR'). Running standalone on the Meta Quest 3 headset, we find the NanoVer iMD-XR client capable of sustaining 36fps for atom counts up to 27k atoms for all three orientations. Fig 7 illustrates the dramatically improved performance that results from using PC-VR with Meta Quest Link, in which case rendering is performed by the laptop GPU and then streamed to the headset by a cable. In such setups, graphics processing is performed on a separate laptop, which can be easily connected to the network by ethernet cable to avoid Wi-Fi bottlenecks. Taken together, these results demonstrate that – for scenarios involving 1 – 4 Meta Quest 3 headsets in an iMD session – the headset's graphics capability limits system size to ~27k atoms (which can be achieved by ensuring the server laptop is cabled to the router). Fig 7 shows that PC-VR enables rendering of up 120k atoms in all 3 user-system orientations, even with a fairly old GeForce GTX 1070. On the latest generation of graphics cards, the final bottleneck is likely to be the speed of the real-time MD simulation itself, which on the modern laptop we tested was limited to 180k atoms.

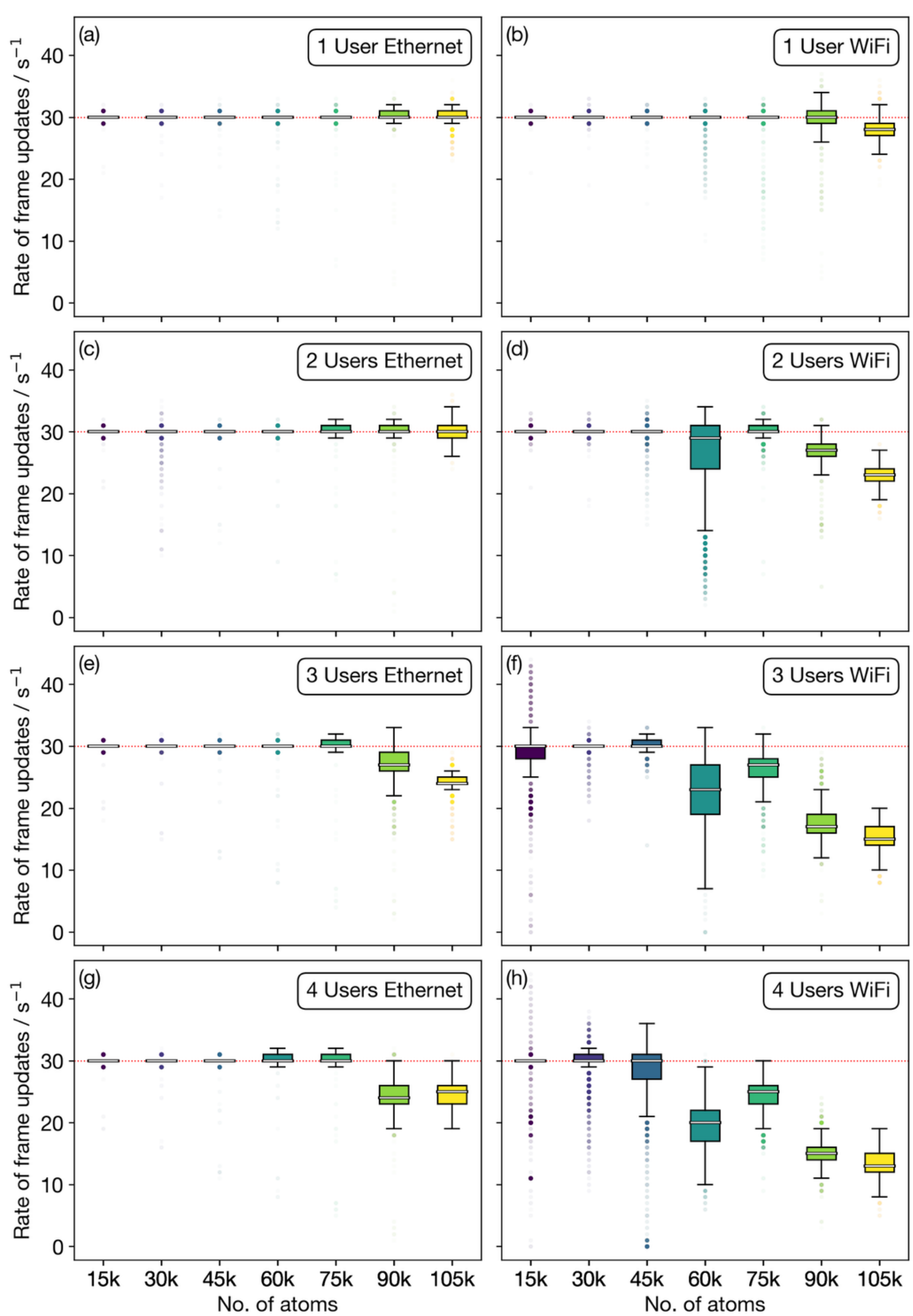

Fig. 6: NanoVer network performance metrics as a function of the number of atoms in the simulation, and the number of iMD-XR users for a setup where the NanoVer server laptop is connected to the router via ethernet cable (panels a, b, c, d) and Wi-Fi (panels e, f, g, h)

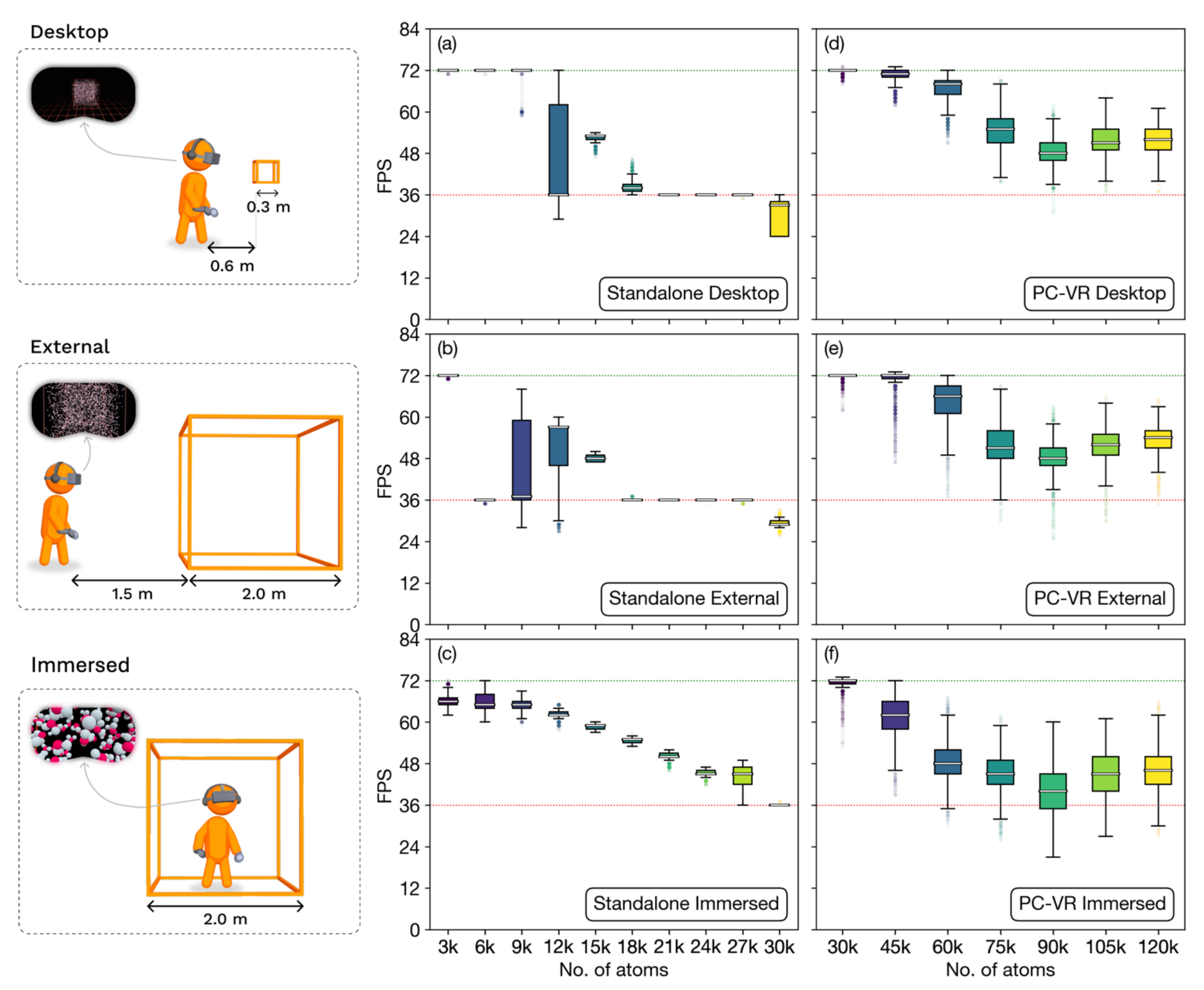

*Fig. 7: NanoVer rendering performance as a function of atom count, measured as* client-side framerate of NanoVer-iMD-XR running on a Meta Quest 3 for the raw "Liquorice" renderer using HMD onboard mobile graphics ('standalone') and a desktop GPU ('PC-VR') across three different POVs: 'Desktop' (where the user is located approx 0.6 m from a water box with dimensions of $(0.3\ \text{m})^3$), 'External' (where the user is located approx 1.5m from a water box with dimensions of $(2.0\ \text{m})^3$), and 'Immersed' (where the user is located inside a water box with dimensions of $(2.0\ \text{m})^3$). Corresponding in-world views for each scenario are shown in the figure.

## 4. Research Applications

In what follows we illustrate the flexibility of the NanoVer software architecture by outlining some of the ongoing applications which are currently underway.

### *4.1 Ligand-Receptor Binding*

iMD-XR enables researchers to use their chemical and spatial intuition to guide a molecular system toward rare macromolecular transitions in high-dimensional systems that would otherwise incur significant computational cost, avoiding the need to define collective variables or reaction coordinates of the sorts that are typically employed in enhanced sampling techniques such as Metadynamics(*95*) or Umbrella Sampling.(*96*) Subsequent analysis on the iMD-XR-sampled trajectories enables the calculation of quantities like work or free energy.(*53*) Ligand-receptor binding is an active application domain for NanoVer: by carrying out real-time molecular manipulations, researchers can gather insight into the molecular interactions and conformational dynamics that impact the process of ligand-receptor binding. Fig 8a for example shows NanoVer being used to study the binding of Tamiflu to neuraminidase, a glycoprotein essential for the influenza life cycle which we have previously investigated using iMD-XR.(*50*) Neuraminidase possesses a flexible loop, whose conformational dynamics appear to influence binding.(*97, 98*) Fig 8b illustrates PatGMac,(*99, 100*) a macrocyclase that produces cyclic peptides, which features a helix-turn-helix loop motif that prevents loss of catalytic activity from the active site being exposed to water. Fig 8c shows the classical serotoninergic psychedelic molecule DMT bound to the HT2A human serotonin receptor embedded in a slab of cell membrane.(*101*) Classical psychedelics molecules like DMT, LSD, and psylocin cause significant subjective effects as a result of interactions with these membrane receptors.(*86, 102*) but assume different binding poses and durations. The receptor lid acts as a latch and constrains the ligand to the binding pocket. Currently we are using iMD-XR to explore how lid motion impacts the conformational binding dynamics for different psychoactive ligands.

Finally, Fig 9 illustrates some preliminary results which we have obtained using iMD-XR to study how the synthetic macroreceptor GluHUT binds to different sugars.(*103, 104*) Specifically, it shows profiles for the cumulative work done as a researcher uses iMD-XR to guide two different sugars out of the GluHUT binding pocket. The work profiles appear to correlate with the experimental binding affinities: glucose has a stronger binding affinity and therefore requires more work to remove from the GluHUT binding pocket, whereas fructose has a weaker binding affinity and therefore requires less work to remove from the GluHUT binding pocket. We are currently working on methods for converting the Fig 9 profiles to free energy curves which can then be compared to results obtained from other free energy binding methods, building on strategies we have previously outlined. (*53*)

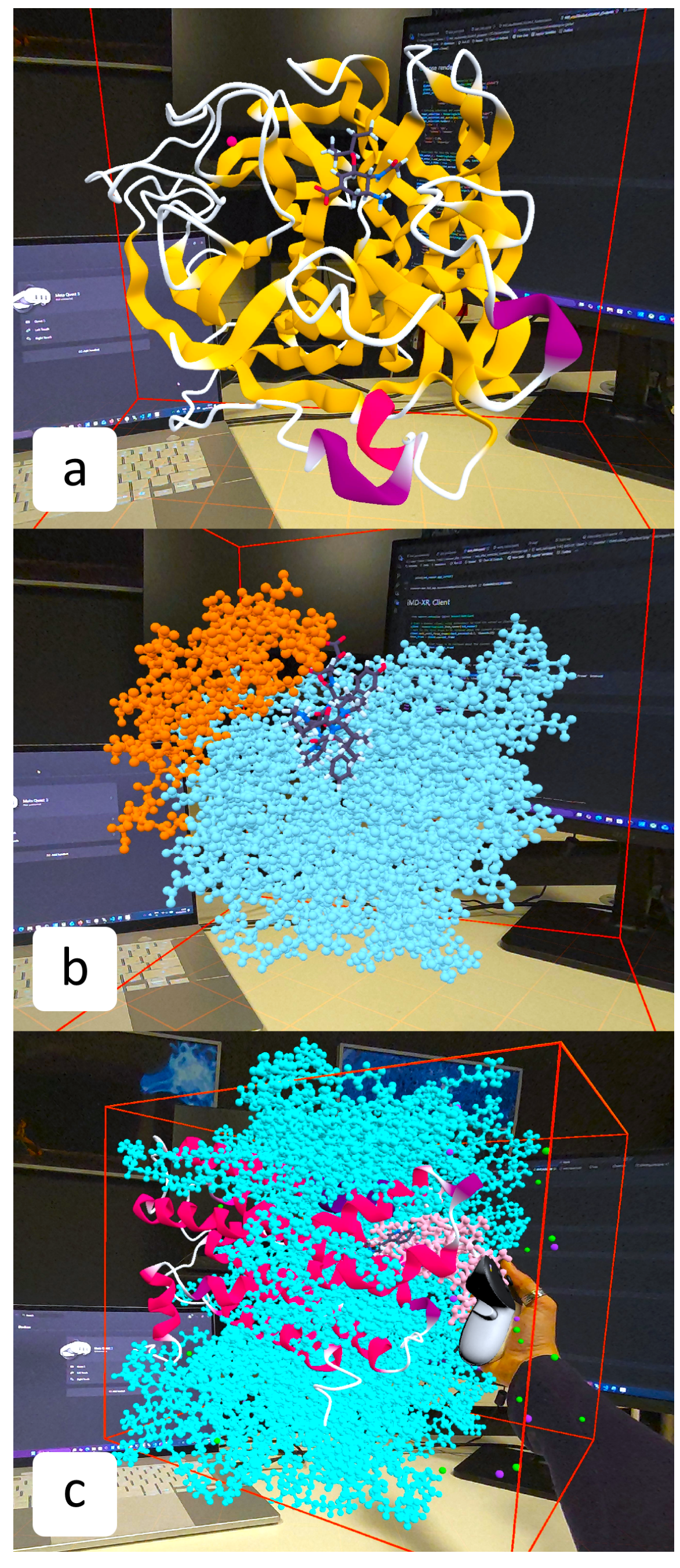

*Fig. 8: first-person POV of a researcher using NanoVer to explore: (a) binding of Tamiflu to neuraminidase; (b) the PatGMac macrocyclase; and (c) DMT bound to the membrane-embedded HT2A human serotonin receptor*

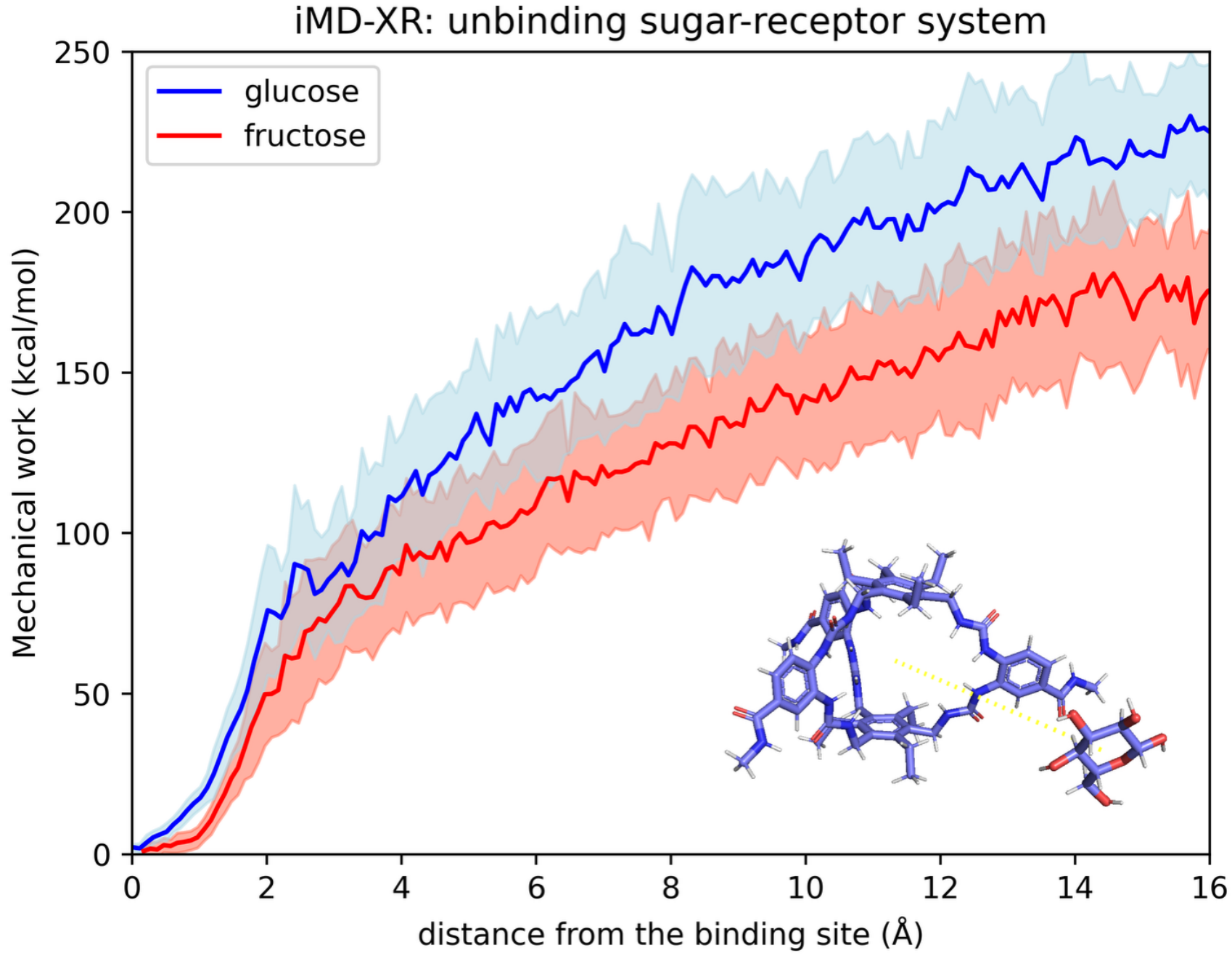

*Fig. 9: Cumulative mechanical work W done of the iMD unbinding trajectories of glucose and fructose from the GLUHUT receptor. The plot shows the average work and the standard deviation as a function of the sugar–binding-site distance. Each sugar was sampled in 25 trajectories.* $W = \sum_{i=2}^{N} F_{user,i-1} \cdot (r_i - r_{i-1})$ *where* $\boldsymbol{F_{user}}$ *is the force applied by the user on the selected atoms,* $\boldsymbol{r}$ *are atoms positions,* $\boldsymbol{i}$ *is the frame index and* $\boldsymbol{N}$ *is the total number of frames.*

### *4.2 iGUESSMD*

As a nonequilibrium enhanced sampling technique, iMD-XR simulations often involve applying large forces to a system over short timescales, driving the system away from equilibrium. Deeks et al(*53*) showed how iMD-XR pathways can be combined with well-established methods like umbrella sampling to recover equilibrium properties like free energies. Other work has illustrated how iMD-XR can be combined with boxed molecular dynamics to recover equilibrium properties.(*105, 106*) These studies suggest that carefully sampled iMD-XR pathways may not be too far from equilibrium, and may thus provide good starting points for quantitative prediction of equilibrium properties. Following this work, we continue to explore new methods for analyzing iMD-XR sampled pathways in order to recover equilibrium properties which can be compared with experiment. To this end, we have recently made progress in developing a workflow that combines iMD-XR with steered molecular dynamics (SMD), a nonequilibrium enhanced sampling technique that (similar to iMD) applies external forces to the system to drive it out of equilibrium and sample rare events of interest. (*107, 108*) Distinct from iMD, SMD generally requires the researcher to define the RC of interest programmatically prior to the simulation, which makes it challenging to explore non-linear processes that involve complex pathways with multiple degrees of freedom. Previous research has demonstrated that SMD simulations can be used to efficiently calculate equilibrium properties (e.g., potentials of mean force) via the Jarzynski equality, (*109*) which relates the external work done on a system during an ensemble of nonequilibrium simulations to the free energy differences between the initial and final states of interest.

Combining the strengths of iMD-XR and SMD, we are testing a method called **i**nteractively **G**enerated **U**ser-defined **E**uclidean **S**tring **S**teered **M**olecular **D**ynamics (iGUESSMD), which is schematically illustrated in Fig. 10. iGUESSMD enables researchers to quickly and intuitively sample iMD-XR pathways and then recover quantitative estimates of equilibrium properties along iMD-XR paths. Rather than abstract the RC into a CV-based space, iGUESSMD defines the reaction coordinate as the path which an atom (or centre of mass of a group of atoms) takes through 3-D space during an iMD-XR simulation. The analytical tools of SMD can then be applied to analyze iMD-XR pathways, for example application of the Jarzynski equality to calculate potentials of mean force (i.e. free energy profiles) along the defined RC. The tools to construct the iGUESSMD workflow are currently being implemented in NanoVer, to be described in detail in a forthcoming publication.

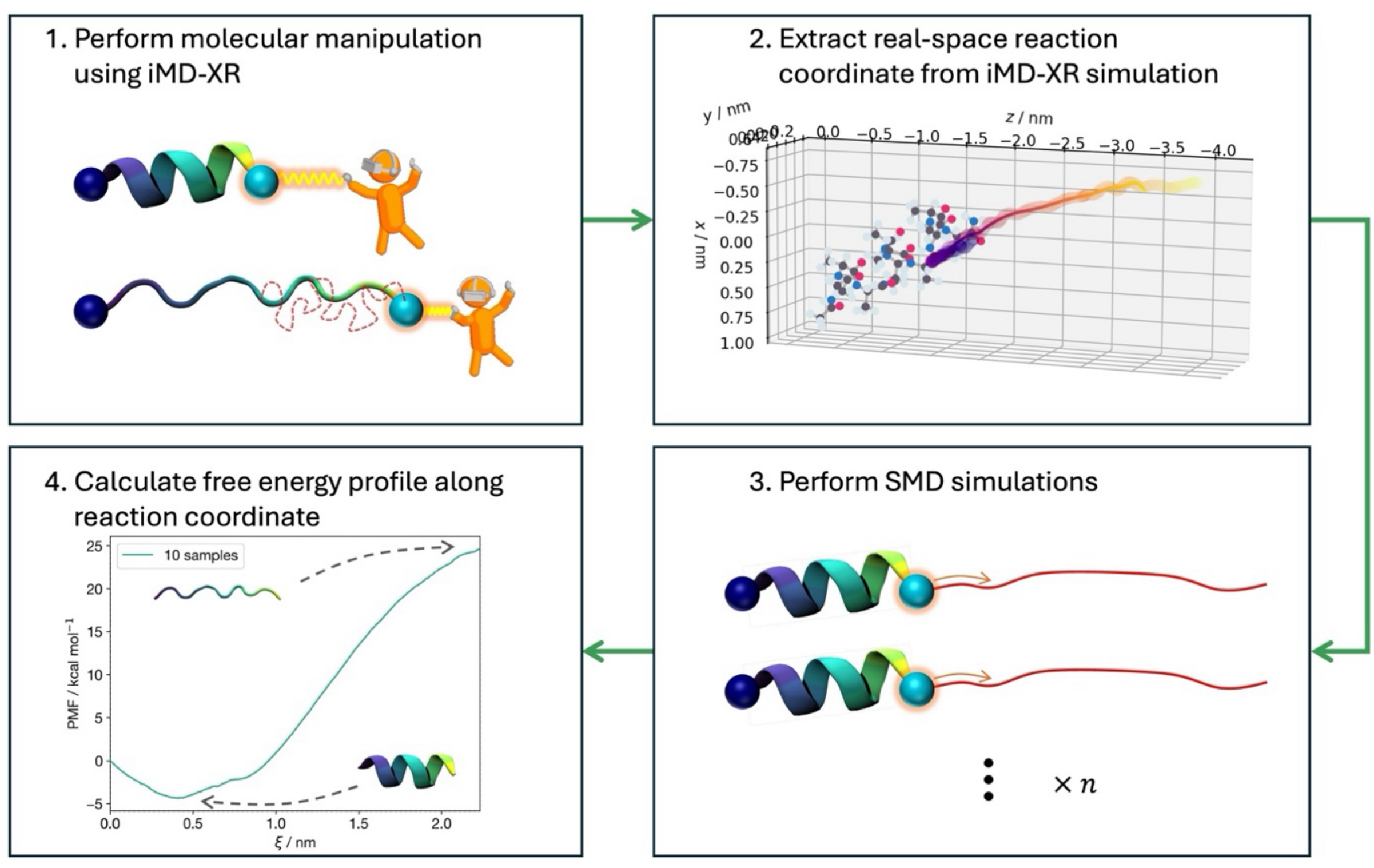

*Fig. 10: Schematic of the iGUESSMD workflow for the helix-to-coil transition of deca-alanine: (1) the researcher performs the molecular manipulation of interest to create the real-space RC; (2) the RC is extracted, interpolated and refined; (3) SMD simulations are performed along the refined RC; and (4) the results of the SMD simulations are used to calculate the free energy profile along the RC.*

### *3.3 Molecular Psychophysics*

Having carried out several demonstrations of NanoVer over the years in both research and outreach contexts, we have noted that participants often comment that they can 'feel' the differences between distinct molecular systems. This is somewhat surprising, given that NanoVer does not include any physical feedback. For example, O'Connor et al. (*48*) documented the so-called "Burke Perception Experiment" (BPE), during which Prof. Kieron Burke remarked that a flexible polypeptide "feels so much different" than a carbon nanotube. To enable more rigorous investigation of this novel perceptual phenomenon and better understand what people mean by 'feeling' molecules, we developed ‘SubtleGame’, a bespoke framework for implementing experimental psychophysics methods within iMD-XR.(*14*) Whereas a typical NanoVer iMD-XR session is relatively unstructured, the SubtleGame XR application leads the player through a series of gamified iMD-XR simulations (including for example tying a knot in a polypeptide as illustrated in Fig. 11). The control logic is managed by a ‘game manager’ (written in Python), which sends commands via the NanoVer server to trigger specific sequences of events in a Unity XR application (e.g., providing in-world menus that lead players onto the next simulation). The NanoVer data streams are recorded to enable data analysis, including the trajectory data, user-applied forces, and user responses.

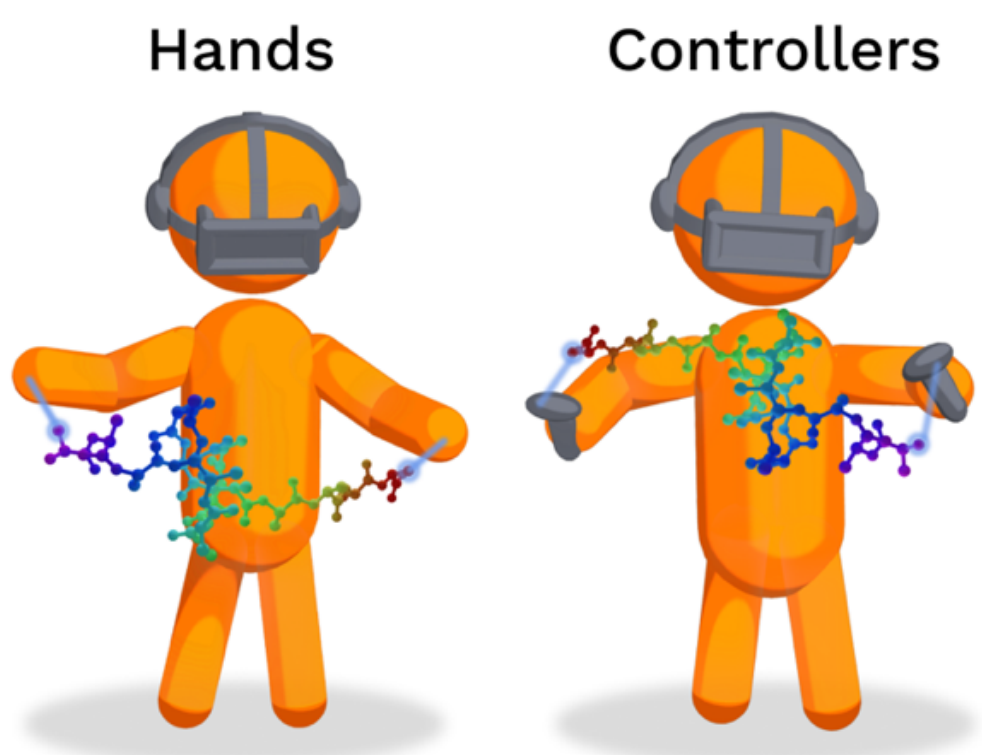

*Fig. 11: Schematic of an iMD-XR user tying a knot in a polypeptide using optical hand-tracking and Handheld XR controllers in SubtleGame.*

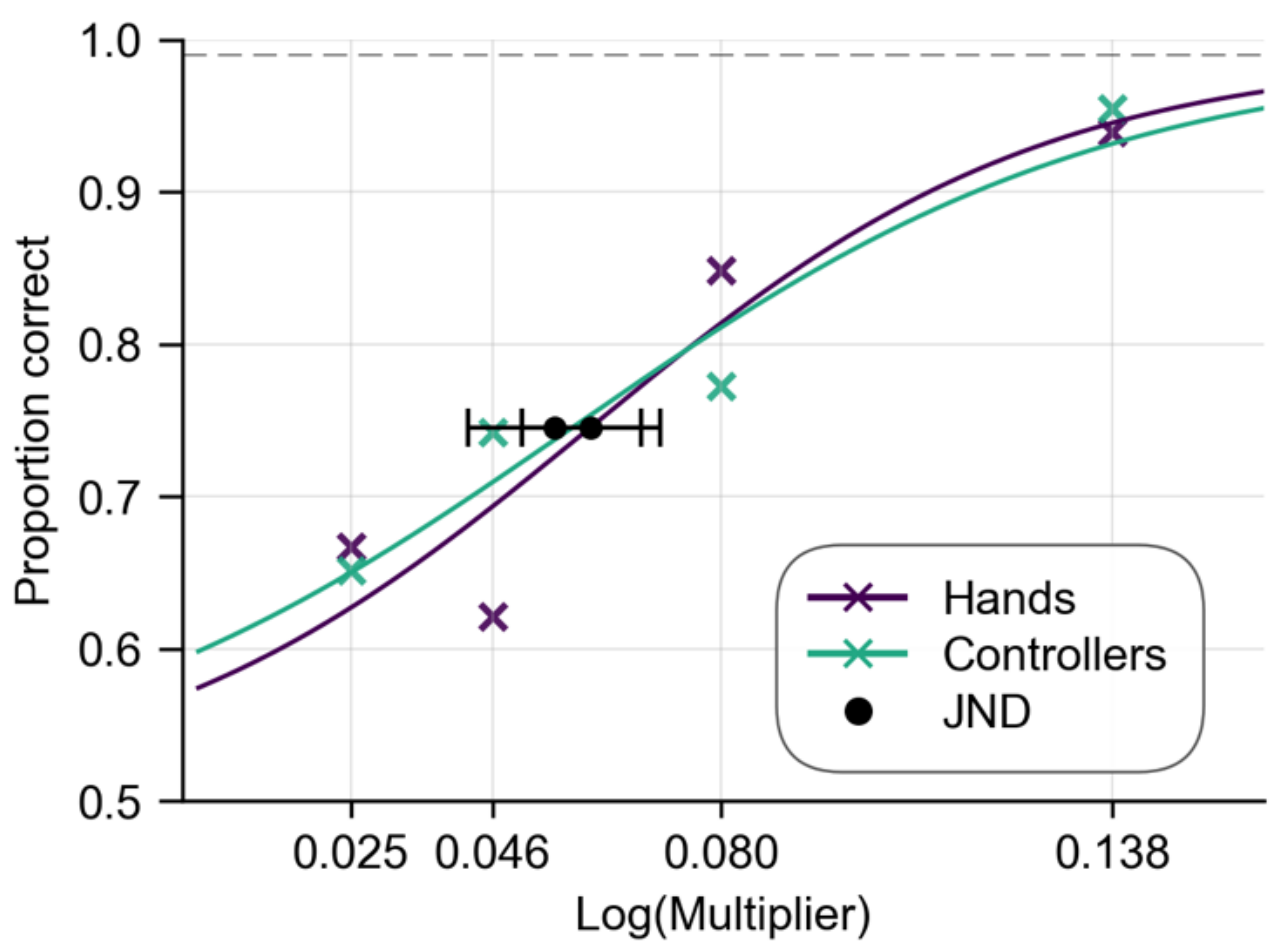

*Fig. 12:* ***Psychometric curves for the Hands and Controllers experimental conditions.*** *The x-axis correlates with the difference in rigidity between the two molecules (with smaller values corresponding to smaller stiffness differences) and the y-axis shows the proportion of correct responses aggregated across participants. The scatter plots give the average probability across the participants for rigidity difference and JND values are given with one standard deviation error bars.*

SubtleGame has enabled us to implement a rigorous two-alternative force-choice (2AFC) experimental psychophysics protocol, which aims to quantify the just-noticeable-difference (JND) at which iMD-XR users can distinguish differences in ‘softness’ or ‘hardness’ of different $C_{60}$ molecules (which is encoded in their corresponding bond angle force constants). This work builds on proof-of-concept results illustrating that such psychophysics approaches could be used to quantify perception of the mechanical properties of molecules in iMD-XR, and that iMD-XR users are able to distinguish bond-angle stiffnesses when interacting with simulations using XR controllers and optical hand tracking (e.g., as illustrated in Fig 11).(*14, 58*) We have since carried out more thorough studies to obtain so-called 'psychometric curves' like that in Fig. 12. Reassuringly, the data appear to follow the sigmoidal form characteristic of psychometric functions: when rigidity differences are small, iMD-XR users are unable to reliably discriminate differences in stiffness, and their performance approaches chance probability (50%). As the difference in stiffness increases, the probability of a correct response increases. The JND is defined as the stimulus magnitude corresponding to the midpoint between chance performance and the asymptotic maximum of the fitted psychometric function. Fig 12 indicates that participants were able to discriminate relatively subtle differences $C_{60}$ rigidity (on the order of ~13-15%) when using handheld controllers or hand-tracking. In ongoing studies, we are investigating the role that interactivity plays in shaping this perception by quantitatively comparing psychometric curves obtained by participants who are able to interact vs. participants that are only able to observe (no interaction). These studies are helping to establish a new domain of 'molecular psychophysics', highlighting molecular simulations as a novel psychophysics paradigm for exploring the detailed mechanisms of human perception in XR environments. Such studies will enable us to quantify the extent to which 'interaction' (the 'i' in iMD-XR) furnishes deeper insight into molecular properties.

***3.4 Training AI agents to sample conformational pathways***

For some iMD-XR applications – e.g., free energy sampling of protein-ligand binding pathways (*53*) – the quality of the results and the corresponding insight depends on the number and quality of iMD-XR trajectories which can be generated, which is limited by human throughput. Given that iMD-XR sessions encode expert spatial intuition regarding molecular structure and function, producing rich time series of atomic positions, velocities, energies, and user-applied forces, we have recently begun to explore a new workflow in which humans train AI agents to sample iMD-XR pathways. Dhouioui et al. recently provided proof-of-principle results illustrating that iMD-XR datasets can be used to train AI agents via imitation learning (IL),(*110*) showing that a convolutional neural network (CNN) trained on 20 expert iMD-XR recordings achieved an $R^2$ of 0.87 for task that involved threading a $CH_4$ molecule through a nanotube. These results illustrate that relatively small expert-generated datasets (which can be obtained in a matter of hours) contain learnable structure that can be used to train a model which achieves good performance. This insight is compatible with recent demonstrations that machine learning can discover reaction coordinates on-the-fly during path sampling with few-shot learning. (*111*)

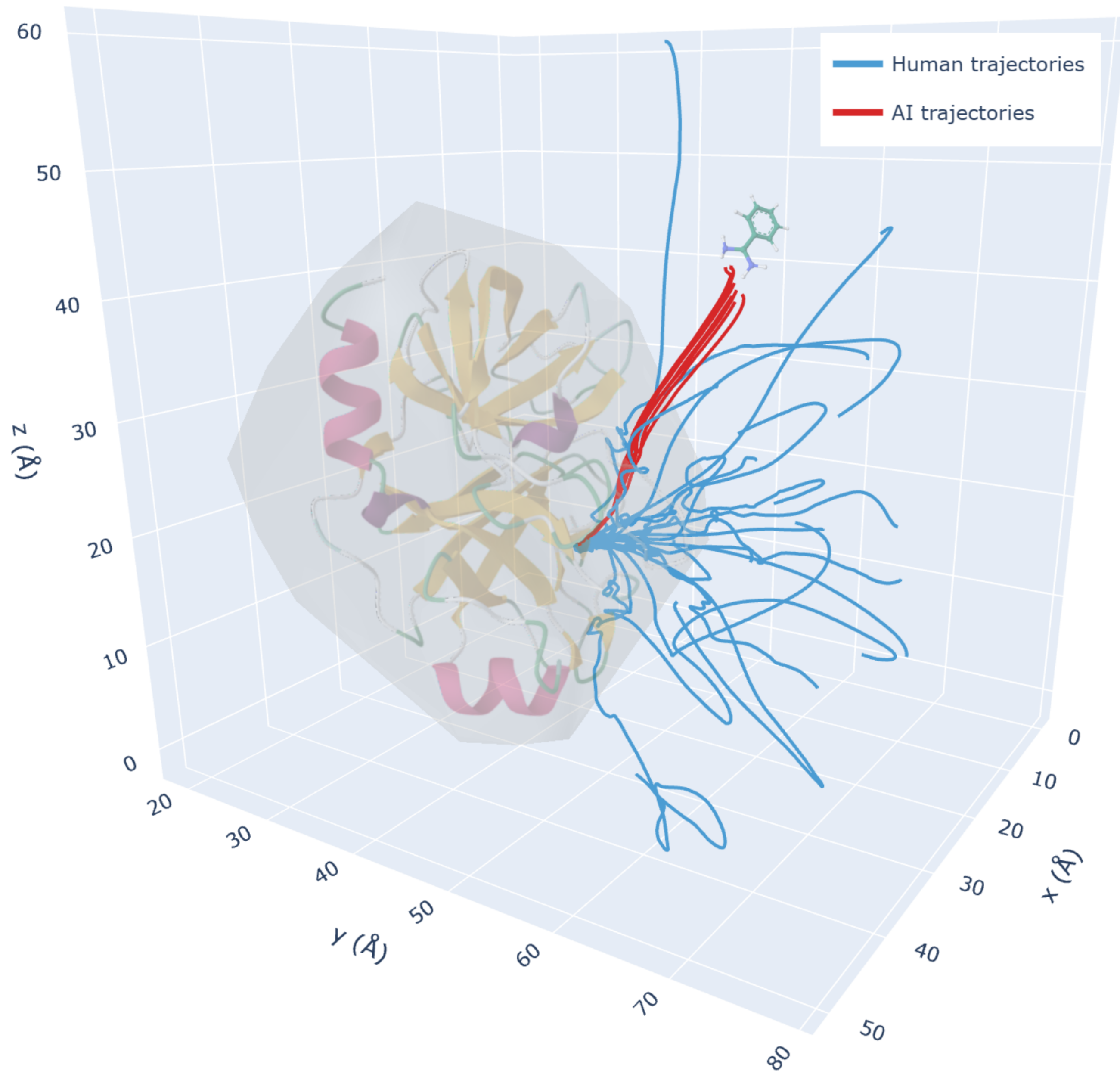

*Fig. 13: Unbinding Trajectories of Benzamidine from Trypsin in NanoVer simulations (human-sampled paths are shown in blue and model-sampled paths in red)*

Recent demonstrations that GPT-inspired models are able to learn complex longer-time dynamics from short unbiased MD simulations(*112*) suggest that the transformer self-attention mechanism captures long-range dependencies across molecular trajectory sequences. As a way of complementing the process of discovery that humans undertake in iMD-XR, we have begun to explore iMD-XR imitation learning using transformer architectures. Specifically, we are exploring a collaborative human-AI discovery workflow in which: (1) human experts generate a modest number of iMD-XR pathways for a conformational transition pathway of interest; (2) these iMD-XR pathways are used to train transformer models via behavioural cloning or adversarial IL; and (3) the resulting AI agents can autonomously drive further exploration, generating an ensemble of new iMD-XR pathways within the hyperdimensional conformational space. Whereas traditional enhanced sampling methods require careful selection of collective variables or reaction coordinates, iMD-XR circumvents this requirement by allowing human research experts to apply forces directly in 3D Cartesian space. AI models trained on the resulting iMD-XR trajectories can then implicitly learn the relevant low-dimensional manifolds governing rare transitions. Fig 13 illustrates some results obtained from applying this methodology to the unbinding of trypsin from benzamidine, a system where we have previously applied iMD-XR.(*53*) While the results are still preliminary, they nevertheless suggest that agents trained using transformer architectures can sample dynamically meaningful iMD-XR pathways with a relatively modest number of expert-sampled pathways.

### *3.5 Path Tinker*

Whereas most computational research tasks involve sitting at a desk, staring at a screen, inspecting 1D text strings or 2D images, typing on keyboard, and using a mouse, sampling an iMD-XR pathway is a more embodied experience, which can be likened to "molecular surgery" insofar as it involves precise movement in 3 dimensions that in some cases requires sustained concentration and motor coordination. Similar to surgery, iMD-XR sampling benefits from slow, thoughtful, and intentional interactions, with steady movements that sometimes require repetition depending on the nuances of the system in question. In fact, surgery remains one of the primary scientific use cases for XR technologies, where it has been used for decades in training,(*15*) with various demonstrated advantages including faster surgical procedures and lower error rates. (*113*) As such, the metaphor of "molecular surgery" to describe iMD-XR sampling has inspired us to consider whether XR methodologies for surgery might be adapted to improve iMD-XR. In particular, we have been investigating whether XR interaction strategies for surgical path planning and navigation (*114*) can be adapted to iMD-XR.

In previously published results where we sampled iMD-XR pathways to subsequently calculate binding free energies in the benzamidine + trypsin system, (*53*) the time required to sample a single iMD-XR pathway was on the order of 10 – 15 minutes. The process of sampling an iMD-XR pathway can be likened to making a 3D 'mid-air' drawing. During this time, the researcher is dynamically guiding the pathway of some component of the molecular system, from a specified initial state to a target end state, based on the real-time visual feedback of the atomic coordinates at time $t$. In recent work, we have begun to explore approaches which preserve the ability of iMD-XR researchers to accurately express their spatial intuition in 3D, but which minimize sources of inaccuracy and fatigue that arise from extended 'mid-air' interactions in XR environments,(*115-117*) which is similarly an important consideration in XR-augmented surgical path planning.

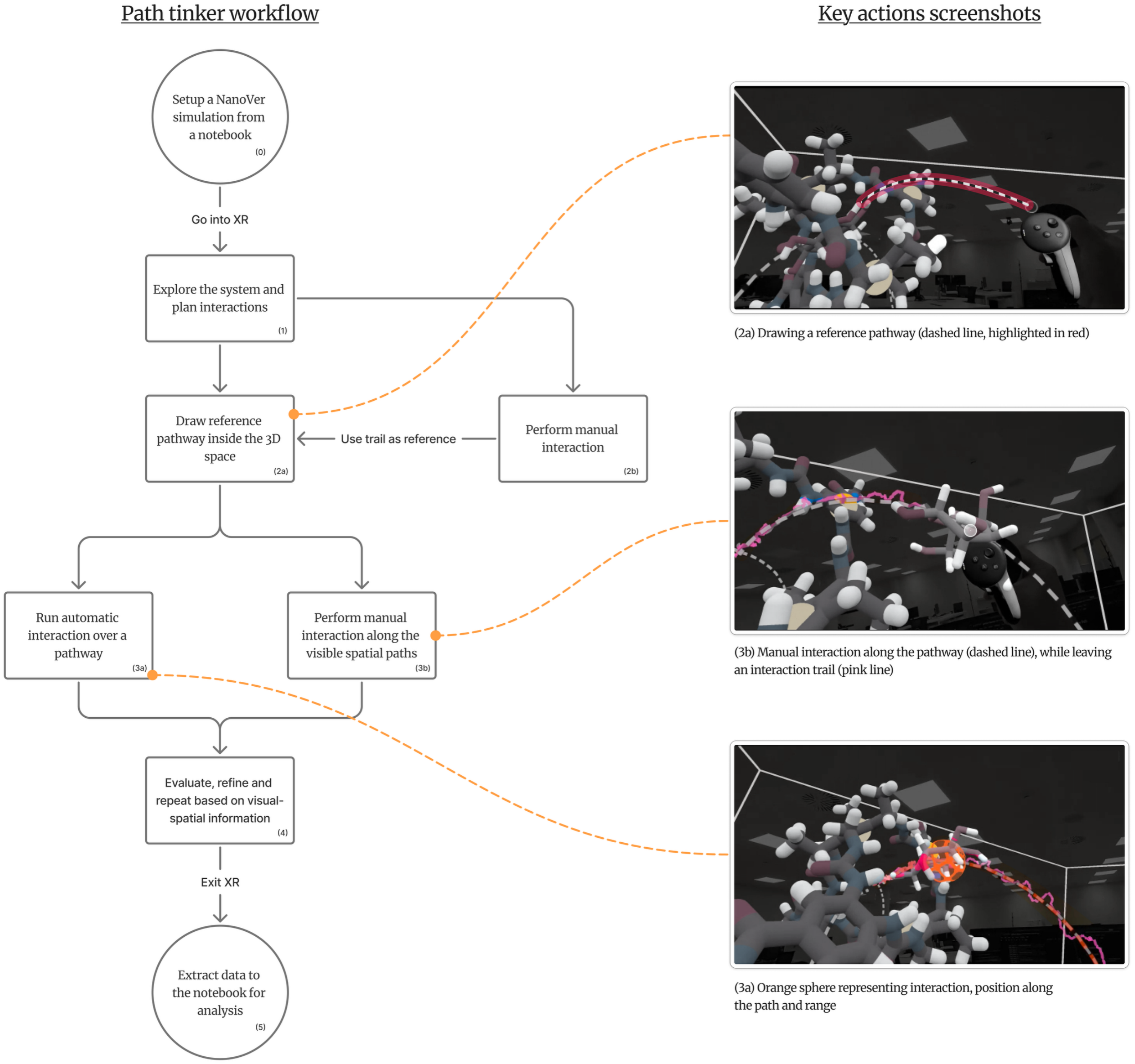

*Fig. 14: Flowchart indicating steps for the "Path tinker" workflow, along with screenshots of key actions (modified to highlight the relevant items). Rectangles represent user actions, arrows show transitions. From setup and systems exploration, to extract pathway data derived from automatic or manual interactions guided by 3D pathway references. Screenshots provided from key steps: (2) Drawing a pathway, (3b) Manual interaction along the pathway, and (3a) Automatic interaction running along the pathway.*

"Path Tinker" is an experimental branch of NanoVer which introduces 3D drawing tools for SMD path planning directly into the 3D XR simulation space. The fundamental design hypothesis guiding Path Tinker is that drawing a pathway in 3D along which force biasing can proceed may be easier and faster versus interactively carrying out the force biasing oneself. The Path Tinker iMD-XR workflow (schematized in Fig 14) involves the following stages: (1) exploration of the molecular system; (2) drawing a 3D iMD-XR pathway; (3) initializing a 'steer agent' to run an SMD simulation along the iMD-XR pathway; (4) inspecting the SMD simulation as it runs; and (5) final evaluation of the results. Building on prior research for improving the accuracy of line drawing in XR

environments,(*118, 119*) Path Tinker implements various drawing tools within the simulation space, for example allowing researchers to set spatial markers, draw 3D pathways, and refine previously drawn paths. Path Tinker's 'steer agents' can automatically apply a force while moving at a constant velocity along a pathway, according to user-defined parameters (force, speed, action range, and coordinate system). Whereas humans have only two hands and are therefore limited to two interactions at a time, Path Tinker agents in principle allow for more interaction sites, allowing the exploration of more complex structural transitions. Path Tinker remains in an experimental stage, and proper evaluations are needed prior to full NanoVer integration. However, preliminary results suggest that (compared to human interactions) the Fig 14 workflow produces work profiles with more stable values, increased coherence across the samples, and requires less time.

## 5. Conclusions & Future Directions

In this article, we have provided an overview of NanoVer, a flexible open-source multi-person framework for iMD-XR, which has been designed for compatibility with recent OpenXR standards, networking protocols, and the latest generation of standalone HMDs. Distinct from its predecessors, NanoVer's 'reality dial' approach enables the user to seamlessly transition between pure virtual reality and augmented reality. NanoVer has been designed to support a variety of different research applications, including for example protein-ligand binding and molecular psychophysics. NanoVer's client/server design enables it to be extended to interface with a range of different analysis tools and force engines. The performance metrics presented in this paper provide a good starting point for evaluating NanoVer's suitability to different systems. Moving forward, we aim to explore how the various research tools being developed for use with NanoVer – iGUESSMD, transformer-trained iMD-XR agents, and Path Tinker – can synergistically interact to support ongoing molecular research applications, and continue to improve the reliability of iMD-XR data for more accurate analysis of molecular processes. We are specifically interested in workflows that combine the strengths of human cognition (high-level critical thinking, design reasoning, and 3D spatial navigation) with the strengths of machine intelligence, creating collaborative cycles of human-machine collaboration. We also hope to explore how these tools can be utilized to construct new immersive educational paradigms, enabling students to better understand the structure and dynamics of the world at the nanoscale. In this regard, we have made progress exploring how approaches being developed in NanoVer can be used in conjunction with WebXR, whose low barrier to entry may enable broader access to XR-based visualization in scientific education, communication, and publishing. Finally, to correspond with the publication of this article, we have recently made a version of NanoVer available on the Meta Horizon store for those with a Meta Quest 3.

## Author Contributions

All authors have contributed to the NanoVer software (either through writing code, carrying out testing and applications, writing documentation, supporting releases, or coordinating team members) and writing the paper (including reading, editing, making figures, providing comments, or coordinating team members). Specific contributions are as follows: MDW coordinated efforts amongst the NanoVer developers, carried out the performance testing, wrote the NanoVer and Performance sections, and coordinated efforts amongst the co-authors; LETC wrote the Path Tinker section and supported app release on the Meta Horizon Store; HJS carried out the performance testing, and wrote the iGUESSMD section; LA wrote the ligand-binding section; MD wrote the AI transformer section; RRW wrote the molecular psychophysics section; DP helped to make figures and videos, especially in relation to the NanoVer Meta Horizon release; SS assisted in coordinating the team of developers and authors; DRG conceptualized the paper, organized the co-authors, and wrote the funding applications that support the work.

## Conflicts of Interest

There are no conflicts to declare.

## Data Availability

The NanoVer iMD-XR client app is available for Meta Quest 3 on Meta Horizon (https://www.meta.com/en-gb/experiences/nanover-imd-xr/33606061302340842/), with build files and source code available on GitHub (https://github.com/IRL2/nanover-imd-xr). The NanoVer python server is available via Conda package (https://anaconda.org/channels/IRL/packages/nanover-server/overview), with tutorial and example notebooks and source code available on GitHub (https://github.com/IRL2/nanover-server-py). The NanoVer documentation is available in the browser (https://irl2.github.io/nanover-docs/), with source code available on GitHub (https://github.com/IRL2/nanover-docs). The notebooks and instrumentation changes used for performance testing are available on GitHub (https://github.com/IRL2/nanover-server-py/tree/tests/performance).

## Acknowledgements

This work is supported by the European Research Council under the European Union's Horizon 2020 research and innovation programme through consolidator Grant NANOVR 866559. D.R.G. recognises the Axencia Galega de Innovación for funding as an Investigador Distinguido through the Oportunius Program, the Xunta de Galicia (Research Center of Galicia accreditation 2024-2027 ED431G-2023/04), and the European Union (European Regional Development Fund - ERDF). We would like to thank all previous contributors to NanoVer. A full list of contributors to NanoVer can be found in the CONTRIBUTORS file in the public repository.

# Supporting Information

## NanoVer: An open-source framework for interactive molecular dynamics in extended reality (iMD-XR) on commodity hardware

Mark D. Wonnacott, Luis Ernesto Toledo Castro, Harry J. Stroud, Ludovica Aisa, Mohamed Dhouioui, Rhoslyn Roebuck Williams, Denis Protopopov, Sila Sobrado, David R. Glowacki*

*Intangible Realities Laboratory (IRL), Centro Singular de Investigación en Tecnoloxías Intelixentes (CiTIUS), University of Santiago de Compostela, 15782, Santiago de Compostela, Spain*
*[*drglowacki@gmail.com](mailto:drglowacki@gmail.com)*

This Supporting information includes the following:

1. A brief description of the version of NanoVer app which is available on the Meta Horizon store (https://www.meta.com/experiences/nanover-imd-xr/33606061302340842/)
2. The corresponding round trip time (RTT) plots for Figures 5 and 6 of the main text

Anybody with a Meta Quest 3 or 3S can download the app to their headset by searching the Meta Horizon Store for 'NanoVer'. Once the app has downloaded, users can either (a) visualize online recordings, or (b) connect to a NanoVer server in order to interact with real-time iMD-XR simulations.

***Visualizing Online Recordings***

To visualize online recordings, select 'Online Demos' in the main menu and then (using the laser pointer to move the scroll bar) choose one of the recordings, which presently include:

a. GluHUT system
b. 17-alanine peptide knot tying
c. Methane threading through a nanotube
d. Buckyballs interaction
e. The Trypsin-Benzamidine protein-ligand system

Compared to the feature set available whilst connected to a NanoVer server instance, the features available while visualizing online recordings are limited to the following:

1. *Play, pause, step through, and reset.* These can be accessed by pressing the 'Menu' button on the left controller.
2. *Modulating the Reality dial.* Press the 'Menu' button on the left controller, select 'Options', then click 'Reality Dial'

To change between recordings, press the 'Menu' button on the left controller, select 'Options', and then 'Quit Simulation'. This will bring you to the 'Online Demos' menu, from which you can select a new system to visualize.

***Connecting to a NanoVer Server***

Interacting with real-time iMD-XR simulations requires setting up a NanoVer server on the same network as that to which the Meta Quest is connected. To do this, follow the instructions in section 2.4 of the main text. Once the server is running and accessible, you can connect to it by selecting 'Local Servers' in the main menu, and then selecting the specific server to which you wish to connect. Once you have connected to the server, you can:

1. *Interact with molecules*. This is done by pressing 'Trigger' buttons on the Meta Quest handsets.
2. *Change the interaction force type.* This is done through the simulation settings menu. Press the 'Menu' button on the left controller and then select a force interaction).
3. *Change the force scale*. This is done by moving the Thumbsticks left/right on the Meta Quest handheld controllers.
4. *Scale, rotate, and translate the simulation bo*x. This is done by pressing both of the 'Grip' buttons on the Meta Quest 3 hand controllers
5. *Change the selection target*. The option to exert a force either on an atom or on the center of mass of an entire residue can be accessed by pressing the 'Menu' button on the left controller.
6. *Play, pause, step through, and reset the simulation.* These can be accessed by pressing the 'Menu' button on the left controller.
7. *Modulate the Reality dial.* Press the 'Menu' button on the left controller, select 'Options', then click 'Toggle PassThrough'

Multi-person colocated setups require a calibration procedure which can be accessed by pressing the 'Menu' button on the left controller, and then selecting 'Options'. As detailed in the main text, interacting with the NanoVer server via Python/Jupyter clients allows the user to control many more aspects of the simulation (including for example visualization styles, selections, etc.) than is possible through the in-world XR interface.

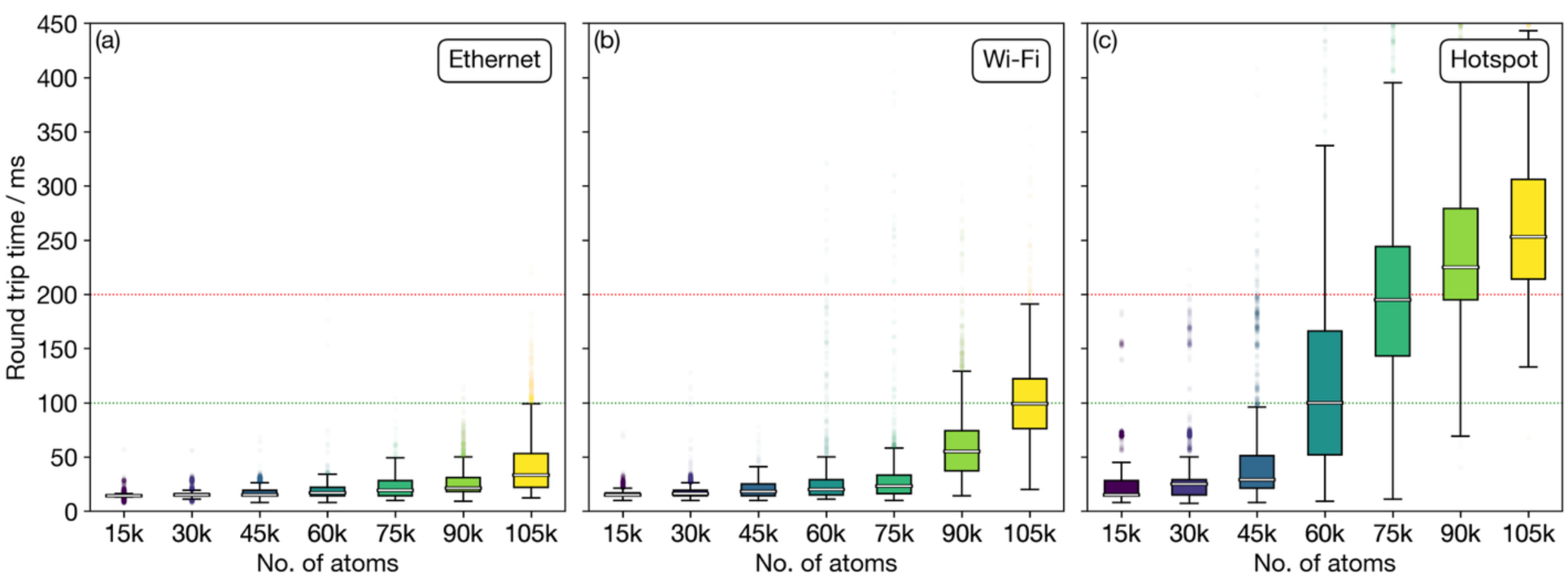

*SI Figure 1: analogue to Fig 5 in the main text, but showing RTT rather than FPS*

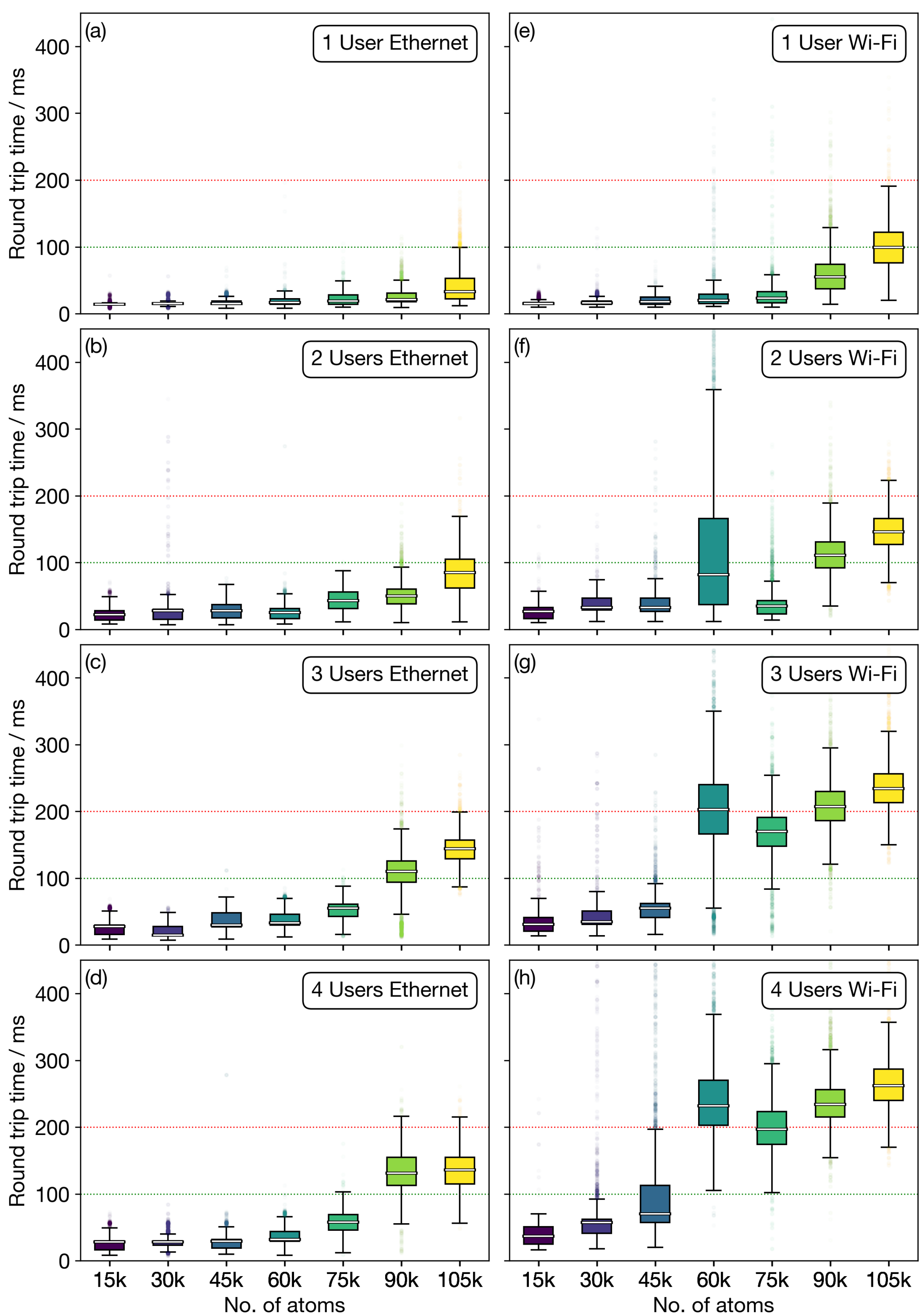

*SI Figure 2: analogue to Fig 6 in the main text, but showing RTT rather than FPS*